\documentclass[aps, preprint, groupedaddress, floatfix, notitlepage]{revtex4-2}
\usepackage{epsfig}
\usepackage{graphicx}
\usepackage{amsmath}
\usepackage{bbold}
\usepackage{tabularx, booktabs}
\usepackage[normalem]{ulem}
\usepackage{xcolor}
\usepackage{multirow}
\usepackage{dutchcal} 
\usepackage[T1]{fontenc}
\usepackage{longtable}
\usepackage{appendix}
\usepackage[colorlinks=true,allcolors=black]{hyperref}

\usepackage{float}
\usepackage[caption = false]{subfig}
\newcolumntype{Y}{>{\centering\arraybackslash}X}
\newcolumntype{Z}{>{\hsize=1.1\hsize\centering\arraybackslash}X}

\begin{document}

\title{
Assessing the feasibility of near-ambient conditions superconductivity in the Lu-N-H system
}

\author{Yue-Wen Fang$^{1,2}$}
\email{yuewen.fang@ehu.eus}
\author{{\DJ}or{\dj}e Dangi{\'c}$^{1,2}$}
\email{dorde.dangic@ehu.es}
\author{Ion Errea$^{1,2,3}$}
\email{ion.errea@ehu.eus}
\affiliation{$^1$ Fisika Aplikatua Saila, Gipuzkoako Ingeniaritza Eskola, University of the Basque Country (UPV/EHU), Europa Plaza 1, 20018 Donostia/San Sebasti{\'a}n, Spain \\
$^2$  Centro de F{\'i}sica de Materiales (CSIC-UPV/EHU), Manuel de Lardizabal Pasealekua 5, 20018 Donostia/San Sebasti{\'a}n, Spain \\
$^3$ Donostia International Physics Center (DIPC), Manuel de Lardizabal Pasealekua 4, 20018 Donostia/San Sebasti{\'a}n, Spain
}
\begin{abstract}
\vspace{1cm}
The recent report of near-ambient superconductivity in nitrogen-doped lutetium hydrides (Lu-N-H) has generated a great interest. However, conflicting results have raised doubts regarding superconductivity. Here, we combine high-throughput crystal structure predictions with a fast predictor of the superconducting critical temperature ($T_c$) to shed light on the properties of Lu-N-H at 1 GPa. None of the predicted structures shows the potential to support high-temperature superconductivity and the inclusion of nitrogen favors the appearance of insulating phases. Despite the lack of near-ambient superconductivity, we consider alternative metastable templates and study their $T_c$ and dynamical stability including quantum anharmonic effects. The cubic Lu$_4$H$_{11}$N exhibits a high $T_c$ of 100 K at 20 GPa, a large increase compared to 30 K obtained in its parent LuH$_3$. Interestingly, it has a similar X-ray pattern to the experimentally observed one. The LaH$_{10}$-like LuH$_{10}$ and CaH$_6$-like LuH$_6$ become high-temperature superconductors at 175 GPa and 100 GPa, with $T_c$ of 286 K and 246 K, respectively. Our findings suggest that high-temperature superconductivity is not possible in stable phases at near-ambient pressure, but metastable high-$T_c$ templates exist at moderate and high pressures. 
\end{abstract}
\maketitle
\newpage

\section{Introduction}

Superconductivity is one of the most fascinating physical properties of matter. Ever since its discovery in mercury below 4.2 K in 1911, humanity has embarked on a restless quest for room-temperature superconductivity at ambient conditions, ceaselessly moving the field forward. Ashcroft suggested that the chemical precompression exerted by the host atoms could boost the superconducting critical temperatures ($T_c$) of hydrogen-rich compounds at lower pressures than pure metallic hydrogen~\cite{H-alloys-Ashcroft-PRL2004}. This idea was further propelled by {\it ab initio} crystal structure prediction techniques at high pressure, which could predict thermodynamically stable 
high-$T_c$ crystal structures~\cite{zhang2017materials,oganov2019structure,pickard2020superconducting,Flores-Livas2020APerspective,Wang2021-HPSTAR-review-hyrides,PengPRL-LaH-PRL,Liu2017-LaH-PNAS}. The main breakthrough arrived in 2015 when Drozdov et al.~\cite{Drozdov2015-H3S-Nature} observed superconductivity at 203 K in H$_3$S at 155 GPa, a compound that had been anticipated theoretically by Duan et al.~\cite{Duan2015-H4S2H2-SciRep}. 
Since then, experiments have reported superconductivity above 200 K in several binary systems such as LaH$_{10}$ ($\sim$250 K at 150 GPa~\cite{LaH10-PRL2019-Somayazulu,Drozdov2019-La-H-Nature}), CaH$_6$ ($\sim$215 K at 172 GPa~\cite{PRL2022-CaH6-exp-YanmingMA}), YH$_9$ ($\sim$243 K at 201 GPa~\cite{Kong2021-NC-Y-H}), and YH$_6$ ($\sim$224 K at 166 GPa~\cite{Kong2021-NC-Y-H,Troyan2021-YH6-AdvMater}). Remarkably, {\it ab initio} calculations support or have even anticipated all these discoveries~\cite{Duan2015-H4S2H2-SciRep,PengPRL-LaH-PRL,Liu2017-LaH-PNAS,Troyan2021-YH6-AdvMater,Errea2015High-Pressure,Errea2016Quantum,Errea2020-LaH10Nature}.

Considering that standard {\it ab initio} crystal structure searches of binary hydrides have been exhausted, the attention is shifting towards ternary hydrides, which offer more possibilities due to the increased complexity of the phase space~\cite{Semenok2021-La-Y-H-Oganov, Wang2021-HPSTAR-review-hyrides}. This recent effort has led to the prediction of ternary superhydrides with high $T_c$s at moderate (below 100 GPa) and even ambient pressures, at which binary superhydrides do not seem to sustain a critical temperature larger than 100 K~\cite{DiCataldo2022-LaBH8-npj,PRB2021-LaBH8-DiCataldo,Metal-borohydrides-DiCataldo-PRB2023-ambient,FrancescoBelli-LaBH8-PRB2022,PRL-hydrides-design-principle-PRL2022,PRL2023-LaBeH8-YanmingMA}. In particular, Dasenbrock-Gammon et al. have recently reported experimental evidence of superconductivity in nitrogen-doped lutetium hydride (Lu-N-H) samples with a room-temperature $T_c$ of 294 K at nearly ambient pressure (1 GPa)~\cite{Dias2023Nature-LuNH}.

This last claim has driven a surge of interest and excitement, but a lively discussion and polemic too. Numerous experimental and theoretical efforts have been made lately trying to replicate or explain the findings uncovered by Dasenbrock-Gammon et al. However, emerging evidence challenges the claim of room-temperature superconductivity. The impressive color change with pressure in their study is suggested to arise from ${Fm\bar{3}m}$ LuH$_2$ due to the presence of an undamped interband plasmon that enters the visible range with increasing pressure and, thus, does not have any impact on superconductivity~\cite{dangic2023-arxiv-color-change-parent-structure}, contrary to the original claim~\cite{Dias2023Nature-LuNH}. Further experimental and theoretical studies~\cite{Shan_2023-CPL-LuH2-colorchange,Ming_2023_Nature_HaihuWen,Kim2023Microscopic,liu2023-arxiv2303.06554,Tao2023-ScienceBulletin-N-LuH2} support that the color change can be explained alone by LuH$_2$, a compound that is known since a long time in which Lu atoms form a face-centered cubic (fcc) lattice and H atoms occupy interstitial tetrahedral sites~\cite{bonnet1977rare-earth}.  
In addition, several experimental and theoretical investigations of the X-ray powder diffraction (XRD) point out that the major peaks in Dasenbrock-Gammon et al.'s study should mostly come from ${Fm\bar{3}m}$ LuH$_2$~\cite{Xie2023_CPL-LuNH,liu2023-arxiv2303.06554,Shan_2023-CPL-LuH2-colorchange,Ming_2023_Nature_HaihuWen,zhang2023-LuH2ArXiv-2303.11063,Zurek2023-LuNH-arXiv2303.15622}. 
These studies have indicated that the parent phase of the Lu-N-H is more likely to be ${Fm\bar{3}m}$ LuH$_2$, rather than the ${Fm\bar{3}m}$ LuH$_3$ as claimed by Dasenbrock-Gammon et al. However, all the existing experimental and theoretical studies~\cite{Daou_1988_LuH2x_EPL,dangic2023-arxiv-color-change-parent-structure,Zurek2023-LuNH-arXiv2303.15622,Tao2023-ScienceBulletin-N-LuH2,MiaoLiu-arxiv-2023epc-LuH} have shown that LuH$_2$ is not superconducting or only shows theoretical $T_c$ on the order of 0.01 K at 0$\sim$1 GPa. {\it Ab initio} crystal structure predictions exploring different stoichiometries of nitrogen-doped lutetium hydrides do not predict any phase with near-ambient superconductivity~\cite{Zurek2023-LuNH-arXiv2303.15622,2023arxiv2304.04447-Lilia-Chris,TianCui-LNH-Matter-Radiation-Extremes2023}.
Furthermore, Ming et al.~\cite{Ming_2023_Nature_HaihuWen} and Cai et al.~\cite{Cai-LuNH-MatterRadiExtremes2023} have successfully obtained nitrogen-doped lutetium hydrides, and asserted that the crystal structures of their samples were the same as those synthesized by Dasenbrock-Gammon et al~\cite{Dias2023Nature-LuNH}. Specifically, the lattice constants of two face-centered-cubic phases (5.03 \AA~and 4.755 \AA) in Cai et al.'s study are in excellent agreement with Dasenbrock-Gammon et al.'s sample A (5.0289 \AA) and sample B (4.7529 \AA).  However, despite this close agreement in lattice constants, neither of the two studies observed superconductivity, even at pressures up to 40 GPa and temperatures as low as 2 K. In addition, Ming et al. have proposed the sample is more appropriately represented as LuH$_{2\pm x}$N$_y$ rather than LuH$_{3 - \delta}$N$_\epsilon$, suggesting that LuH$_2$ is more likely to be the parent phase.  It should be noted that the crystal structure of ${Fm\bar{3}m}$ LuH$_3$ is identical to ${Fm\bar{3}m}$ LuH$_2$ with an extra hydrogen atom located at the octahedral interstitial site~\cite{dangic2023-arxiv-color-change-parent-structure}.

The controversial results require a comprehensive understanding of the properties of Lu-N-H system.  Herein, we report high-throughput crystal structure calculations in the Lu-H and Lu-N-H systems and screen the potential $T_c$ of the predicted compounds with a simple descriptor based on electronic properties~\cite{Belli2021-network-value-NC}. 
Since LuH$_2$ and LuH$_3$ have been suggested to be the potential parent phases of the near-ambient superconducting Lu-N-H, 
we focus the structural search on derivations of both of them with and without nitrogen.
By screening more than 15,000 structures at 1 GPa, our prediction has resulted in the discovery of 638 phases located within 0.24 eV per atom above the convex hull, a reasonable limit for the synthesizability of metastable phases, with 214 of them metallic. 
Our results suggest that these 1-GPa-phases are improbable to manifest high-temperature superconductivity, as deduced by predicting their  $T_c$s with the networking value model~\cite{Belli2021-network-value-NC}. 
Our findings suggest that, when seeking to develop metallic lutetium hydrides at 1 GPa, doping with nitrogen should be avoided as its large electronegativity removes electrons from hydrogen sites and promotes insulating phases.  
Moreover, we identify tens of stable metallic phases that exhibit XRD features strongly resembling the experimental XRD, suggesting that many of the predicted structures have a Lu arrangement not far from the fcc lattice and that H or N atoms occupy interstitial sites. Considering that XRD is not capable of distinguishing them, one should approach XRD structural assignments in the Lu-H-N system with care. As a result of the absence of high-temperature superconductivity at 1 GPa, we study the dynamical stability and superconducting properties of high-symmetry lutetium hydrides with and without nitrogen at higher pressures in crystal structures that favor high $T_c$s. We find that quantum anharmonic effects foster dynamical stability at lower pressures in all cases and strongly impact the phonon spectra. Specifically, cubic Lu$_4$H$_{11}$N exhibits a high $T_c$ of 100 K at a moderate pressure of 20 GPa. Upon increasing pressure, CaH$_6$-like $Im\bar{3}m$ LuH$_6$ and LaH$_{10}$-like ${Fm\bar{3}m}$ LuH$_{10}$ are found to maintain high $T_c$s of 246 K and 289 K, respectively, at 100 GPa and 175 GPa.

\section{Results \& Discussions}
\subsection{Phase diagram} 

\begin{figure*}[h]
\includegraphics[angle=0,width=0.7\textwidth]{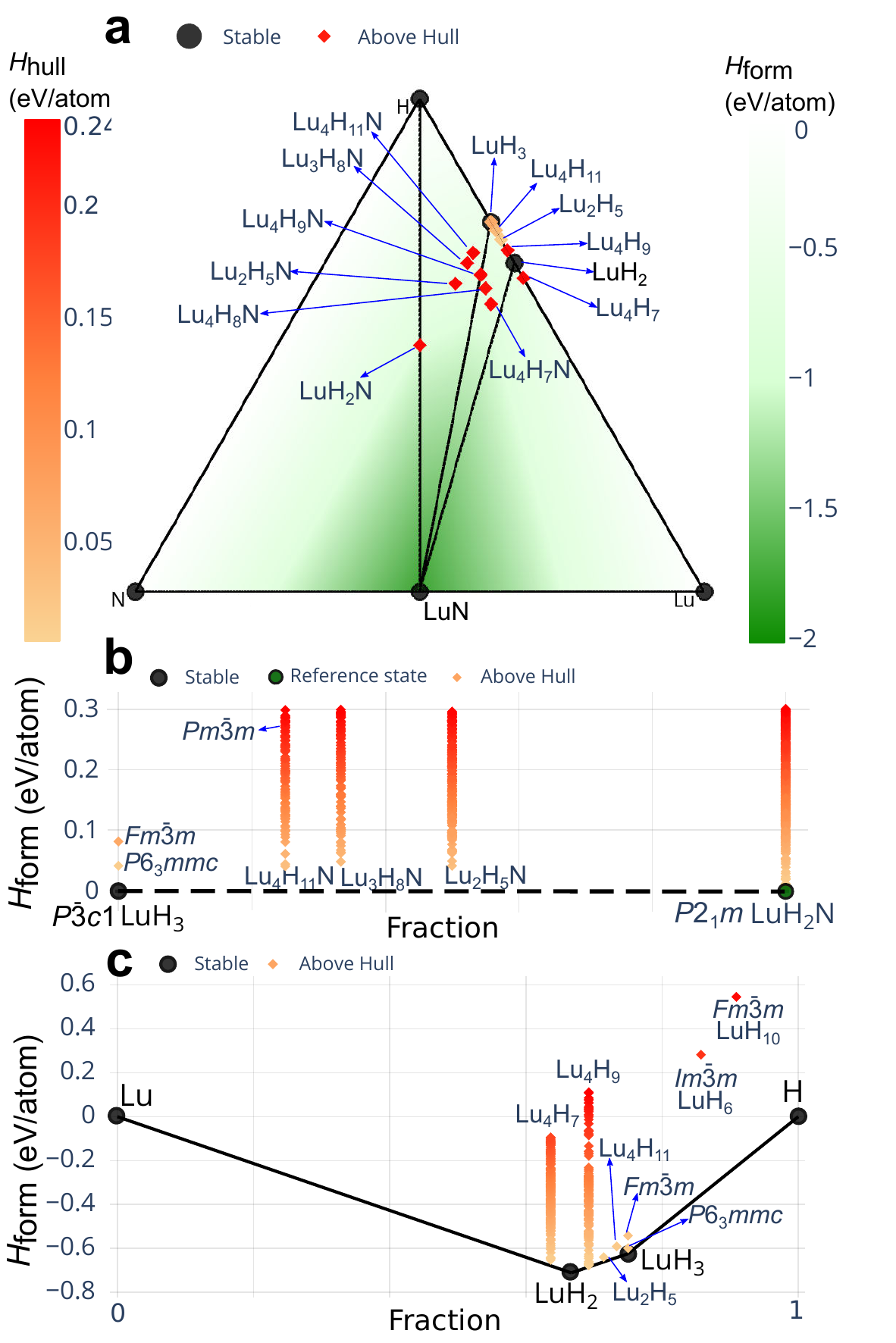}
\caption{\label{fig:luhn_phase_diagram}  
\textbf{The phase diagram of the Lu-N-H system.} 
\textbf{a} The phase diagram of Lu-N-H with respect to elements, in which compounds with ${H_{\rm hull}}$ $\leq$ 0.24 eV${\cdot}$atom$^{-1}$ are shown. \textbf{b} The phase diagram relative to stable ${P\bar{3}c1}$ LuH$_3$ (black circle) and metastable ${P2_1/m}$ LuH$_2$N  (red circle) , in which ${H_{\rm hull}}$ $\leq$ 0.3 eV${\cdot}$atom$^{-1}$.  \textbf{c} The phase diagram of Lu-H with respect to elements, in which ${H_{\rm hull}}$ $\leq$ 0.8 eV${\cdot}$atom$^{-1}$. The black circles in all panels represent the thermodynamically stable phases. Square markers refer to phases located above the convex hull. The line between LuH$_3$ and LuH$_2$ in panel (\textbf{c}) is set to dash because it is not a convex hull as those in other panels.
}
\end{figure*}

The phase diagram for the Lu-N-H system at 1 GPa is constructed by the convex hull in 
Figure~\ref{fig:luhn_phase_diagram}(a), in which the stable and possibly metastable phases up to an enthalpy  of 0.24 eV${\cdot}$atom$^{-1}$ above the convex hull (abbreviated as ${H_{\rm hull}}$ $\leq$ 0.24 eV${\cdot}$atom$^{-1}$) are shown by circles and squares, respectively. The enthalpy calculations are performed without considering the zero-point ionic energy, i.e., considering just the Born-Oppenheimer energy. The phase diagram includes the known stable binary phases ${P\bar{3}c1}$ LuH$_3$, ${Fm\bar{3}m}$ LuN, ${Fm\bar{3}m}$ LuH$_2$, and multiple artificially constructed structures (e.g. Lu$_2$H$_5$) based on ${Fm\bar{3}m}$ LuH$_2$ and LuH$_3$ by adding/removing H atoms at tetragonal/octahedral sites. In addition, the phase diagram comprises 638 phases predicted through high-throughput crystal structure screening over around 15,000 crystal structures. Among the 638 predicted structures within this enthalpy cutoff, there are 214 metallic and 424 insulating phases. 

The ${P2_1/m}$ LuH$_2$N, with structure ID in our database of 2fu\_LuH2N\_389, is identified as the lowest-enthalpy state among the ternary Lu-H-N compounds at 1 GPa. 
Dynamical stability in the harmonic approximation of the ${P2_1/m}$ LuH$_2$N is examined by performing finite displacement DFT calculations of 2$\times$2$\times$2 supercells. The phonon spectra and phonon density of states (DOS), as depicted in Fig.~\ref{fig:2fu_LuH2N_389_phonon_supercell222}, demonstrate the absence of any imaginary modes, thereby substantiating the dynamic stability of the system. Seeing that the ${P2_1/m}$ LuH$_2$N is found to be thermodynamically metastable near the convex hull, it is likely to be synthesized under proper experimental conditions. However, the electronic DOS of the ${P2_1/m}$ LuH$_2$N as shown in Fig.~\ref{fig:2fu_LuH2N_389_e_dos} exhibits a band gap of $\sim$ 2 eV, undoubtedly excluding it to be a superconducting phase. Because ${P2_1/m}$  LuH$_2$N is the lowest-enthalpy state and is dynamically stable, the phase diagram with respect to stable ${P\bar{3}c1}$ LuH$_3$ and metastable LuH$_2$N is explicitly shown in Fig.~\ref{fig:luhn_phase_diagram}(b), in which the structures up to ${H_{\rm hull}}$ of 0.3 eV${\cdot}$atom$^{-1}$ are displayed.

Fig.~\ref{fig:luhn_phase_diagram}(c) shows the phase diagram with respect to elements, in which the binary hydrides with ${H_{\rm hull}}$ $\leq$ 0.8 eV${\cdot}$atom$^{-1}$ are included.  The ${Fm\bar{3}m}$ phase of LuH$_3$, which has been proposed by Dasenbrock-Gammon et al.~\cite{Dias2023Nature-LuNH} as the parent compound of the room-temperature superconductor, is however found to be located above the convex hull by around 82 meV${\cdot}$atom$^{-1}$. This enthalpy difference at this pressure is not expected to be overcome by ionic zero-point energy even if quantum anharmonic effects are considered~\cite{Errea2020-LaH10Nature}. 

To identify potential hosts for superconductivity, we have focused our attention on the stability of the 214 metallic phases with ${H_{\rm hull}}$ $\leq$ 0.24 eV${\cdot}$atom$^{-1}$ that have been found in our high-throughput crystal structure predictions. In order to assess the phonon stability of the 214 metallic phases in the harmonic approximation, we have performed high-throughput DFT calculations over 165,000 supercells generated by the finite displacement method (see Methods). These calculations identify 57 dynamically stable metallic phases including 20 Lu$_4$H$_7$, 19 Lu$_4$H$_9$, 11 Lu$_4$H$_7$N, 2 Lu$_4$H$_9$N, 2 Lu$_4$H$_8$N, 2 Lu$_3$H$_8$N, and 
1 Lu$_4$H$_{11}$N. The harmonic phonon spectra of 20 Lu$_4$H$_7$ and 19 Lu$_4$H$_9$ are displayed in Fig.~\ref{fig:metallic-phonopy-stable-Lu4H7} and Fig.~\ref{fig:metallic-phonopy-stable-Lu4H9}, respectively. Additionally, the phonon spectra of the 18 ternary Lu-N-H phases are presented in Fig.~\ref{fig:metallic-phonopy-stable-LuHN}. The dynamical stability of these 57 phases is well evidenced by the depicted phonon spectra, which do not show imaginary phonon modes. 

The ratio of the contribution of hydrogen to the total electronic density of states (DOS) at the Fermi level is widely suggested to be an important descriptor of  superconductivity in superhydrides. Therefore, the hydrogen fraction of the total DOS at the Fermi level, a.k.a. ${H_{\rm DOS}}$, are computed for the 214 metallic phases. Table~\ref{table:57metallic-Tc-XRD} lists ${H_{\rm DOS}}$, space group symbols, and the enthalpy distances above the convex hull (${H_{\rm hull}}$) of the 57 stable metallic phases at 1 GPa. A complete list of the 214 metallic phases regardless of the dynamical stability is available in Table~\ref{table:214metallic-Tc-XRD}. As shown in Table~\ref{table:57metallic-Tc-XRD}, 55 out of the 57 stable metallic phases show very low ${H_{\rm DOS}}$ ranging from 0.02 to 0.05 (or 2$\%$ to 5$\%$), and only the Lu$_4$H$_9$N (ID: 1fu\_Lu4H9N\_136) and Lu$_3$H$_8$N (ID: 1fu\_Lu3H8N\_216) with the same space group ${P3m1}$ show large ${H_{\rm DOS}}$ exceeding 20$\%$. The low ${H_{\rm DOS}}$ widely observed in the predicted structures signals the low critical temperature or even the absence of superconductivity of the predicted phases. 

To gain more insights into the possible onset of superconductivity among the metallic phases, an understanding beyond  ${H_{\rm DOS}}$ is necessary. However, performing electron-phonon coupling calculations for all these low-symmetry systems is not feasible. As an alternative, Belli et al. have proposed that a physical quantity termed as the networking value ($\phi$), which is the electron localization function value that creates an isosurface spanning throught the whole crystal, can be used to predict easily $T_c$. In fact, $\phi$ exhibits a stronger correlation with the actual $T_c$ of hydrogen-based superconductors than any other descriptor used so far, and can be used to predict $T_c$ with an accuracy of about 60 K with the formula ${T_c = (750\phi H_f {H_{\rm DOS}}^{-3} - 85)}$ K, where $H_f$ is the hydrogen fraction in the compound~\cite{Belli2021-network-value-NC}. 
The results of  $\phi$ and the estimated $T_c$ of the dynamically stable phases and all the 214 metallic states irrespective of the stability are included in Table~\ref{table:57metallic-Tc-XRD} and Table~\ref{table:214metallic-Tc-XRD}, respectively (an Excel file is also available in Data Availability). Our results indicate that the upper limit of the predicted $T_c$ among the 214 metallic states is only ${13.94 \pm 60~\text{K}}$. These results collectively indicate that the metallic phases predicted from the high-throughput crystal structure prediction are unlikely to host high-temperature superconductivity at 1 GPa.

\begin{center}
\captionsetup{width=\linewidth}
\begin{longtable*}{@{\extracolsep{\fill}}ccccccc}
\caption{\label{table:57metallic-Tc-XRD}
The 57 dynamically stable metallic states at 1 GPa. The superconducting transition temperatures ($T_c$) are estimated by using the networking value model in Ref.~\cite{Belli2021-network-value-NC}. Only structures below 0.24 eV${\cdot}$atom$^{-1}$ above the convex hull are considered. ${H_{\rm hull}}$ represents the enthalpy distance above convex hull. ID refers to the unique identifier of a structure entry in our database. ${H_{\rm DOS}}$ is the hydrogen fraction of the total density of states at the Fermi level, $\phi$ refers to the networking value.  XRD similarity refers to the degree of similarity in the simulated X-ray diffraction (XRD) patterns between the structure being analyzed and the reference structure (i.e. $Fm\bar{3}m$ LuH$_2$).
} 
\\
\hline
ID & $H_{\rm hull}$ (eV$\cdot$atom$^{-1}$) & Space group & ${H_{\rm DOS}}$ & $\phi$ & $T_c$ ($\pm$60 K)  & XRD similarity ($\%$) \\
\hline
\endfirsthead
\hline
ID & $H_{\rm hull}$ (eV$\cdot$atom$^{-1}$) & Space group & ${H_{\rm DOS}}$ & $\phi$ & $T_c$ ($\pm$60 K)   & XRD similarity ($\%$) \\
\hline
\endhead
\hline
\multicolumn{6}{r}{\textit{Continued on next page}} \\
\endfoot
\hline
\endlastfoot

1fu\_Lu3H8N\_216 & 0.24 & $P\bar{3}m1$ & 0.26 & 0.31 & 13.94 & 75.03 \\
1fu\_Lu4H7\_495 & 0.17 & $R3m$ & 0.04 & 0.42 & -17.88 & 64.48 \\
2fu\_Lu4H7\_242 & 0.17 & $Cmc2_1$ & 0.04 & 0.40 & -19.35 & 12.78 \\
2fu\_Lu4H9\_373 & 0.05 & $Pba2$ & 0.03 & 0.39 & -19.56 & 93.61 \\
2fu\_Lu4H9\_34 & 0.02 & $Cm$ & 0.04 & 0.33 & -25.32 & 98.53 \\
1fu\_Lu4H9\_213 & 0.01 & $Cmmm$ & 0.04 & 0.34 & -25.58 & 99.07 \\
2fu\_Lu4H9\_231 & 0.01 & $P2_1/m$ & 0.04 & 0.34 & -26.0 & 99.48 \\
1fu\_Lu4H9\_190 & 0.11 & $R3m$ & 0.05 & 0.30 & -28.28 & 62.05 \\
2fu\_Lu4H9\_12 & 0.15 & $Cm$ & 0.03 & 0.32 & -31.37 & 94.71 \\
2fu\_Lu4H7\_51 & 0.19 & $P4/mmm$ & 0.02 & 0.43 & -32.23 & 68.09 \\
2fu\_Lu4H9\_233 & 0.01 & $Cmmm$ & 0.04 & 0.30 & -32.3 & 98.97 \\
1fu\_Lu4H9\_139 & 0.06 & $Pmmm$ & 0.03 & 0.32 & -32.68 & 98.66 \\
1fu\_Lu4H9\_175 & 0.01 & $Pm\bar{3}m$ & 0.03 & 0.32 & -32.85 & 99.12 \\
2fu\_Lu4H7\_391 & 0.04 & $Pc$ & 0.03 & 0.35 & -33.55 & 97.01 \\
2fu\_Lu4H7\_62 & 0.08 & $Fmm2$ & 0.03 & 0.35 & -33.57 & 98.19 \\
1fu\_Lu4H9\_182 & 0.01 & $Pmmm$ & 0.03 & 0.31 & -34.12 & 98.91 \\
2fu\_Lu4H7\_149 & 0.09 & $P2_1/m$ & 0.03 & 0.33 & -34.5 & 98.41 \\
2fu\_Lu4H7\_66 & 0.03 & $Cmcm$ & 0.03 & 0.34 & -34.7 & 97.52 \\
1fu\_Lu4H7\_318 & 0.12 & $Amm2$ & 0.03 & 0.34 & -36.45 & 97.97 \\
1fu\_Lu4H7\_478 & 0.04 & $Amm2$ & 0.03 & 0.34 & -36.53 & 99.15 \\
1fu\_Lu4H7\_417 & 0.03 & $P\bar{4}3m$ & 0.02 & 0.35 & -36.77 & 97.87 \\
2fu\_Lu4H7\_375 & 0.09 & $Cmc2_1$ & 0.03 & 0.33 & -37.66 & 95.62 \\
1fu\_Lu4H7\_64 & 0.07 & $Cm$ & 0.03 & 0.31 & -37.98 & 97.45 \\
1fu\_Lu4H7\_288 & 0.16 & $Amm2$ & 0.02 & 0.35 & -38.25 & 65.62 \\
1fu\_Lu4H7\_467 & 0.23 & $P\bar{4}m2$ & 0.02 & 0.39 & -38.88 & 59.39 \\
2fu\_Lu4H9\_401 & 0.02 & $P4/mmm$ & 0.04 & 0.26 & -39.37 & 99.12 \\
2fu\_Lu4H7\_63 & 0.17 & $Cmc2_1$ & 0.03 & 0.32 & -39.55 & 62.64 \\
1fu\_Lu4H7\_29 & 0.16 & $P1$ & 0.02 & 0.33 & -40.67 & 68.36 \\
1fu\_Lu4H7N\_358 & 0.10 & $R3m$ & 0.02 & 0.36 & -41.32 & 68.35 \\
2fu\_Lu4H9\_225 & 0.20 & $P1$ & 0.03 & 0.26 & -41.51 & 97.85 \\
2fu\_Lu4H9\_184 & 0.10 & $I4mm$ & 0.02 & 0.28 & -42.75 & 22.18 \\
1fu\_Lu4H7\_81 & 0.17 & $Cm$ & 0.02 & 0.31 & -44.59 & 57.65 \\
1fu\_Lu4H9\_167 & 0.01 & $I4/mmm$ & 0.03 & 0.25 & -44.64 & 98.97 \\
1fu\_Lu4H8N\_5 & 0.07 & $R\bar{3}m$ & 0.03 & 0.27 & -44.97 & 51.47 \\
1fu\_Lu4H7N\_471 & 0.15 & $Pmm2$ & 0.03 & 0.28 & -45.06 & 67.44 \\
1fu\_Lu4H7\_44 & 0.19 & $C2$ & 0.02 & 0.30 & -45.25 & 58.91 \\
2fu\_Lu4H9\_111 & 0.10 & $Cc$ & 0.03 & 0.24 & -45.43 & 91.48 \\
1fu\_Lu4H7N\_166 & 0.15 & $P1$ & 0.04 & 0.26 & -45.95 & 27.14 \\
1fu\_Lu4H9N\_10 & 0.21 & $P3m1$ & 0.04 & 0.24 & -46.57 & 37.17 \\
1fu\_Lu4H9\_6 & 0.16 & $P1$ & 0.04 & 0.22 & -46.88 & 64.86 \\
2fu\_Lu3H8N\_116 & 0.20 & $P1$ & 0.03 & 0.24 & -47.27 & 49.77 \\
1fu\_Lu4H9N\_136 & 0.04 & $P3m1$ & 0.31 & 0.11 & -48.99 & 13.12 \\
1fu\_Lu4H11N\_251 & 0.19 & $Cm$ & 0.05 & 0.19 & -49.69 & 66.42 \\
1fu\_Lu4H7\_378 & 0.14 & $Cmm2$ & 0.02 & 0.27 & -50.04 & 79.79 \\
1fu\_Lu4H7N\_11 & 0.11 & $Cm$ & 0.03 & 0.24 & -52.56 & 91.97 \\
1fu\_Lu4H7\_379 & 0.20 & $Pm$ & 0.02 & 0.22 & -55.31 & 88.22 \\
1fu\_Lu4H8N\_105 & 0.07 & $Pm$ & 0.03 & 0.21 & -56.18 & 74.46 \\
1fu\_Lu4H7N\_377 & 0.15 & $Cm$ & 0.03 & 0.20 & -56.47 & 21.50 \\
1fu\_Lu4H7N\_491 & 0.14 & $P1$ & 0.03 & 0.21 & -57.5 & 84.14 \\
1fu\_Lu4H9\_129 & 0.15 & $Imm2$ & 0.03 & 0.17 & -57.64 & 14.80 \\
1fu\_Lu4H7N\_172 & 0.17 & $P1$ & 0.03 & 0.21 & -57.7 & 83.10 \\
2fu\_Lu4H9\_67 & 0.14 & $P4_2/nmc$ & 0.02 & 0.20 & -58.06 & 40.45 \\
1fu\_Lu4H7N\_488 & 0.17 & $Cm$ & 0.02 & 0.22 & -58.37 & 67.76 \\
1fu\_Lu4H7N\_372 & 0.15 & $P4mm$ & 0.03 & 0.19 & -60.64 & 52.78 \\
2fu\_Lu4H9\_420 & 0.04 & $Cmcm$ & 0.03 & 0.15 & -62.03 & 93.12 \\
1fu\_Lu4H7N\_494 & 0.15 & $Cm$ & 0.02 & 0.16 & -64.6 & 63.68 \\
1fu\_Lu4H7N\_67 & 0.02 & $R\bar{3}m$ & 0.03 & 0.06 & -76.61 & 53.46 \\
\end{longtable*}
\end{center}

\begin{figure}[ht]
\includegraphics[angle=0,width=0.9\textwidth]{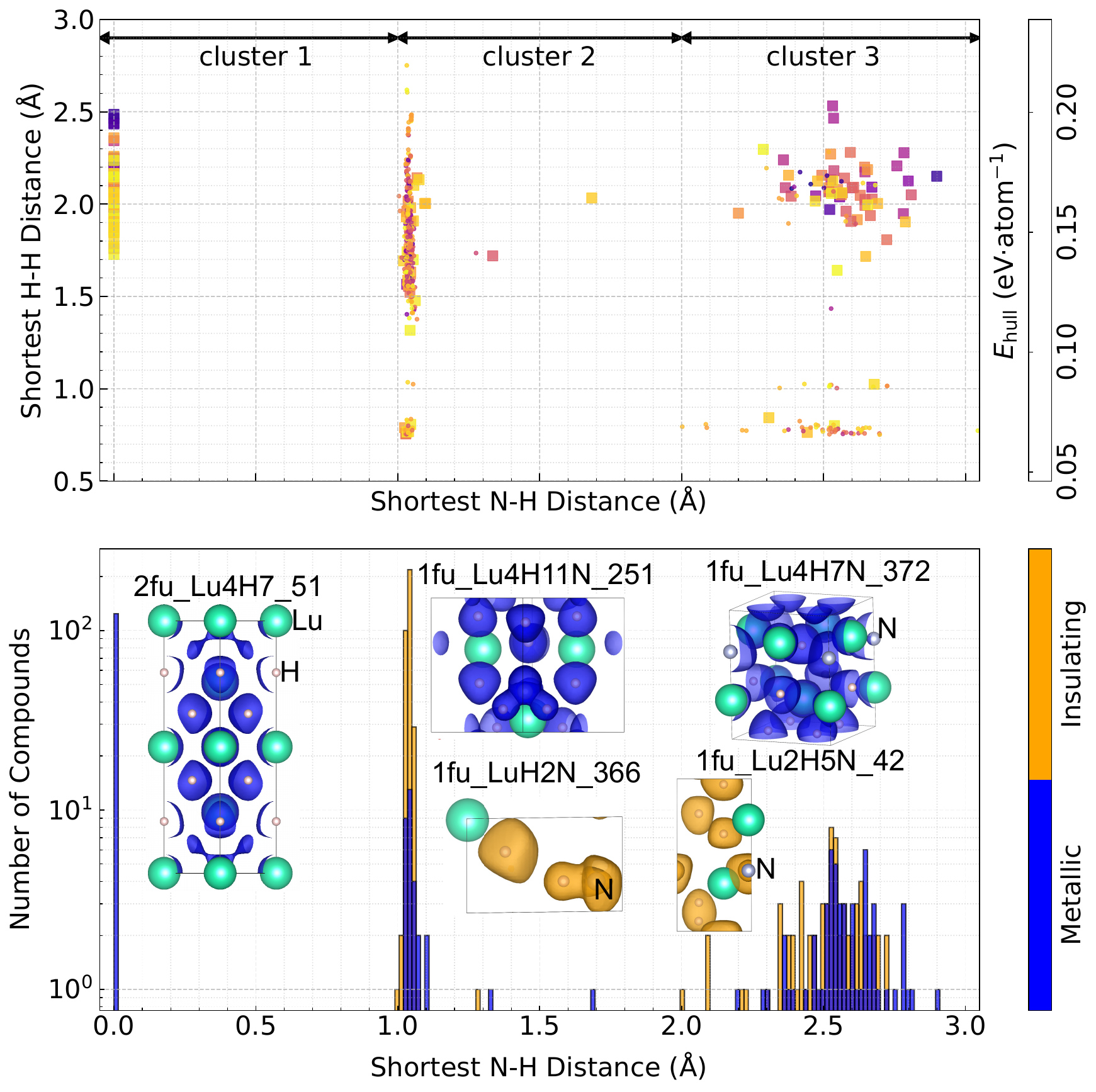}
\caption{\label{fig:HH-NH-distances}  
\textbf{The analysis of shortest N-H and H-H distances.} 
Upper panel: The comparison between minimal distances of N-H and H-H  for 124 binary Lu-H systems and 514 ternary Lu-N-H systems. The dots and squares refer to insulating and metallic states, respectively. The color bar shows the enthalpy distance above the convex hull from 0 to 0.24 eV${\cdot}$atom$^{-1}$. 
Lower panel: The number of compounds distributed with the shortest N-H distance. The inset shows the electron localization functions of 5 representative compounds.
The isosurface values of 2fu\_Lu4H7\_51, 1fu\_Lu4H11N\_251, 1fu\_LuH2N\_366, 1fu\_Lu4H7N\_372, and 1fu\_Lu2H5N\_42 are
set to 0.52, 0.75, 0.75, 0.52, and 0.75, respectively. The blue (orange) bars/isovalues indicate the metallic (insulating) nature.}
\end{figure}

\subsection{The interatomic distances and Bader charge analysis}

In the upper panel of Fig.~\ref{fig:HH-NH-distances}, we show the shortest N-H and H-H distances for the 638 Lu-H and Lu-N-H compounds within an enthalpy  of 0.24 eV${\cdot}$atom$^{-1}$ above the convex hull (i.e. ${H_{\rm hull}}$ $\leq$ 0.24 eV${\cdot}$atom$^{-1}$) irrespective of their dynamical stability. For the 124 binary Lu-H systems where N is absent, the shortest N-H is artificially set to 0.0 \AA. As we can see, all the binary hydrides are metallic and show an H-H distance from 1.75 to 2.5 \AA. At the same pressure, the H-H distance in the thermodynamically stable ${Fm\bar{3}m}$ LuH$_2$, where all H atoms occupy the tetrahedral site (H$_{\rm tetrahedral}$) is 2.48 \AA. In this crystal, the distance between a tetrahedral and an octahedral site is 2.15 \AA. This suggests that the hydrogen atoms in most of the binary predicted hydrides are close to a tetrahedral or octahedral arrangement in an fcc lattice.
The majority of the 90 ternary metallic Lu-N-H compounds, 64 out of 90, also exhibit H-H distances larger than 1.75 \AA. 
Observing all the previously documented superhydrides compiled in Ref.~\cite{Belli2021-network-value-NC}, it can be seen that none of them exhibit $T_c$ $>$ 50 K when their shortest H-H distance exceeds 1.75 \AA. This analysis further indicates that neither the binary nor ternary metallic phases predicted from our high-throughput crystal structure prediction is likely to host high-temperature superconductivity. 

Based on the shortest N-H distance, these 638 compounds can be classified into three distinct clusters, which are indicated by double-headed arrows in Fig.~\ref{fig:HH-NH-distances}.  Cluster 1, which only contains binary Lu-H compounds, consists of 124 phases that are all metallic. Cluster 2 comprises 352 insulating phases and 32 metallic phases, characterized by N-H distances $<$ 2 \AA~and $\le$ 1 \AA. Cluster 3 is composed of a total of 130 compounds whose shortest N-H distances are all larger than 2 \AA, with 72 of them being insulating and 58 of them being metallic.
The histogram in the lower panel of Fig.~\ref{fig:HH-NH-distances}, in which the vertical axis uses a logarithmic scale, explicitly displays the distribution of the number of compounds based on their shortest N-H distance.
In contrast to cluster 1 comprising 124 binary Lu-H compounds, which exhibit exclusively metallic behavior in the absence of N, the ternary compounds in cluster 2 and cluster 3 containing N exhibit altered characteristics across a total of 514 compounds. The presence of N is found to lead to a significant proportion of 424 insulating phases, which accounts for approximately 82.5$\%$ of all ternary compounds. This notable proportion underscores the crucial role of N in promoting insulating behavior rather than metallic behavior within the Lu-N-H system.

In cluster 1 consisting of 124 Lu-H compounds, it is observed that all the shortest H-H distances span the range of 1.75 \AA~to 2.80 \AA. The presence of relatively large H-H distances prevents the formation of H$_2$ molecules or H-H chains, which typically contribute to forming the insulating states. The observation of relatively large H-H distances provides a clear explanation for the exclusively metallic behavior of the binary compounds in cluster 1. However, as noted in Tables~\ref{table:57metallic-Tc-XRD} and ~\ref{table:214metallic-Tc-XRD}, the contribution of hydrogen to the electrons at the Fermi level is minor in most of these compounds, where Lu $d$ states dominate at the Fermi level. An illustrative example is shown in the inset of the lower panel of Fig.~\ref{fig:HH-NH-distances}, depicting the electron localization function at the isosurface of 0.52 in the 2fu\_Lu4H7\_51 structure. In this binary Lu$_4$H$_7$, a distinct separation exists between the hydrogen atoms, characterized by the shortest H-H distance of 2.18 \AA. Furthermore, the bonding observed between hydrogen and lutetium is ionic.

Among the 514 ternary compounds in clusters 2 and 3, cluster 2 accounts for approximately 74.7$\%$ (384 compounds) of all ternary Lu-N-H systems. The compounds in cluster 2 are distinguished by their small N-H distances and a noticeable propensity towards insulating character. It is found that 99.5$\%$ of the compounds in cluster 2, including 352 insulating entries and 30 metallic entries, feature a shortest N-H distance of approximately $\sim$ 1.1 \AA. This particularly short N-H distance corresponds to the strong covalent bonding between N and H, which is evidenced by the electron localization functions of metallic 1fu\_LuH2N\_366 and insulating 1fu\_Lu4H11N\_251 belonging to cluster 2 in the lower panel of Fig.~\ref{fig:HH-NH-distances}. Only two phases in this cluster have a shortest N-H distance larger than 1.1 \AA, and both of them are metallic. It should be noted that the N-H bond length of NH$_3$ molecule is generally between 1.0 and 1.1 \AA, and purely ammonia is perfectly insulating. Furthermore, by surveying all the compounds exclusively composed of N and H in Materials Project~\cite{Jain2013-MaterProject} that are located within 0.2 eV$\cdot$atom$^{-1}$ above the convex hull, these N-H compounds all show insulating properties and the shortest N-H bond lengths are all in the range of 1.0 $\sim$ 1.1 \AA. Thus, our finding suggests that the presence of N strongly favors insulating phases due to the formation of strongly covalent bonds between N and H atoms. The strong covalent N-H bonds result both in NH or NH2 units in most Lu-N-H structures predicted in our work.

Compared to the small magnitude of the average shortest N-H distance in cluster 2, the shortest N-H distance of cluster 3 is more than twice the former, spanning from 2.0 to 3.05 \AA. There are 72 insulators and 58 metals in cluster 3. In addition, the upper panel of Fig.~\ref{fig:HH-NH-distances} demonstrates a strong correlation 
between the electronic properties and the shortest H-H distance. Among the 71 compounds with shortest H-H distances larger than 1.75 \AA~belonging to cluster 3, 52 compounds ($73.2\%$) are metallic and 19 compounds are insulating. Analogous to the binary Lu-H compounds in cluster 1, long H-H distances favor the presence of metallic states dominated by electrons coming from Lu. The example structure with ID 1fu\_Lu4H7N\_372, hosting 
the shortest N-H distance of 2.59 \AA~and the shortest H-H distance of 2.28 \AA, shows ionic features of N and H ions. 
On the contrary, there are 53 insulators and 6 metals whose shortest H-H distances are smaller than 1.75 \AA~in this cluster.
In particular, 52 of these 53 insulating compounds have the shortest H-H distances below 1.1 \AA, indicating that the formation of H-H molecules plays a crucial role in determining their insulating character. To exemplify this, the electron localization isosurface with the value of 0.75 of the 1fu\_Lu2H5N\_42 structure is shown in the lower panel of Fig.~\ref{fig:HH-NH-distances}, which unambiguously displays the presence of an H-H molecule with an H-H distance of 0.78 \AA.
These analyses of the interatomic distances in Clusters 2 and 3 imply that metallic phases are favored when the shortest H-H distance exceeds 1.75 \AA~and the shortest N-H distance surpasses 2.0 \AA. However, the ternary Lu-N-H compounds, unfortunately, do not satisfy the aforementioned conditions in most cases, thereby resulting in the promotion of insulating character.

To better understand the electronic properties of the 638 Lu-H and Lu-N-H compounds with ${H_{\rm hull}}$ $\leq$ 0.24 eV${\cdot}$atom$^{-1}$, we have performed a Bader charge analysis for all these compounds. The average Bader charges of H and N for each compound are explicitly shown in Fig.~\ref{fig:bader-charge} versus the shortest N-H distance and shortest H-H distance. 
Among the 514 ternary compounds, the average Bader charge of N in metallic phases and insulating phases is -1.70$e$ and -1.64$e$, respectively. It implies that N in metallic phases obtains a slightly greater number of electrons than in insulating cases. However, this difference is not significant, which can also be seen in panels (a) and (c) of Fig.~\ref{fig:bader-charge}. In contrast to the small fluctuation of the average Bader charge of N, the average Bader charge of H in metallic phases is around -0.59$e$, which is nearly twice the value of insulating phases (-0.32$e$). This difference reveals that gaining more electrons at the hydrogen site is crucial for entering the metallic state. The Bader charge of hydrogen versus the shortest N-H distance, as exemplified in panel (b) in Fig.~\ref{fig:bader-charge}, demonstrates that longer N-H distances can be advantageous for hydrogen atoms to gain additional electrons.
In other words, the presence of nitrogen does not enhance the ability of hydrogen to acquire electrons. Thus, it is comprehensible why all the 124 binary Lu-H cases without N, in which the average Bader charge of H is -0.71$e$, manifest metallic states. 
Therefore, the charge around H atoms plays a more significant role than around nitrogen in determining electronic properties. 
Panels (b) and (d) in Fig.~\ref{fig:bader-charge} demonstrate that, among all metallic compounds, a greater acquisition of electrons occurs primarily in compounds where the shortest N-H distance is around 2.5 \AA, while simultaneously ensuring that the shortest H-H distance is greater than 1.75 \AA. This further suggests that, when aiming to develop metallic states in lutetium hydrides, it is advisable to avoid or keep nitrogen away from the hydrogen site. This allows H to acquire more electrons from Lu, promoting the formation of metallic states. Based on statistical data on all previously reported superconducting hydrides in Ref.~\cite{Belli2021-network-value-NC}, it is found that all superconducting critical temperatures in the literature are lower than 50 K if their shortest H-H distances $\ge$ 1.75 \AA. In our case, by examining all 214 metallic states, 188 of them show the shortest H-H distances $\ge$ 1.75 \AA. This also implies that it is unlikely to find high-temperature superconductivity in the structures from the high-throughput structure screening at 1 GPa.

\subsection{XRD comparison at 1 GPa}\label{subsec:xrd}

\begin{figure}[ht]
\includegraphics[angle=0,width=0.99\textwidth]{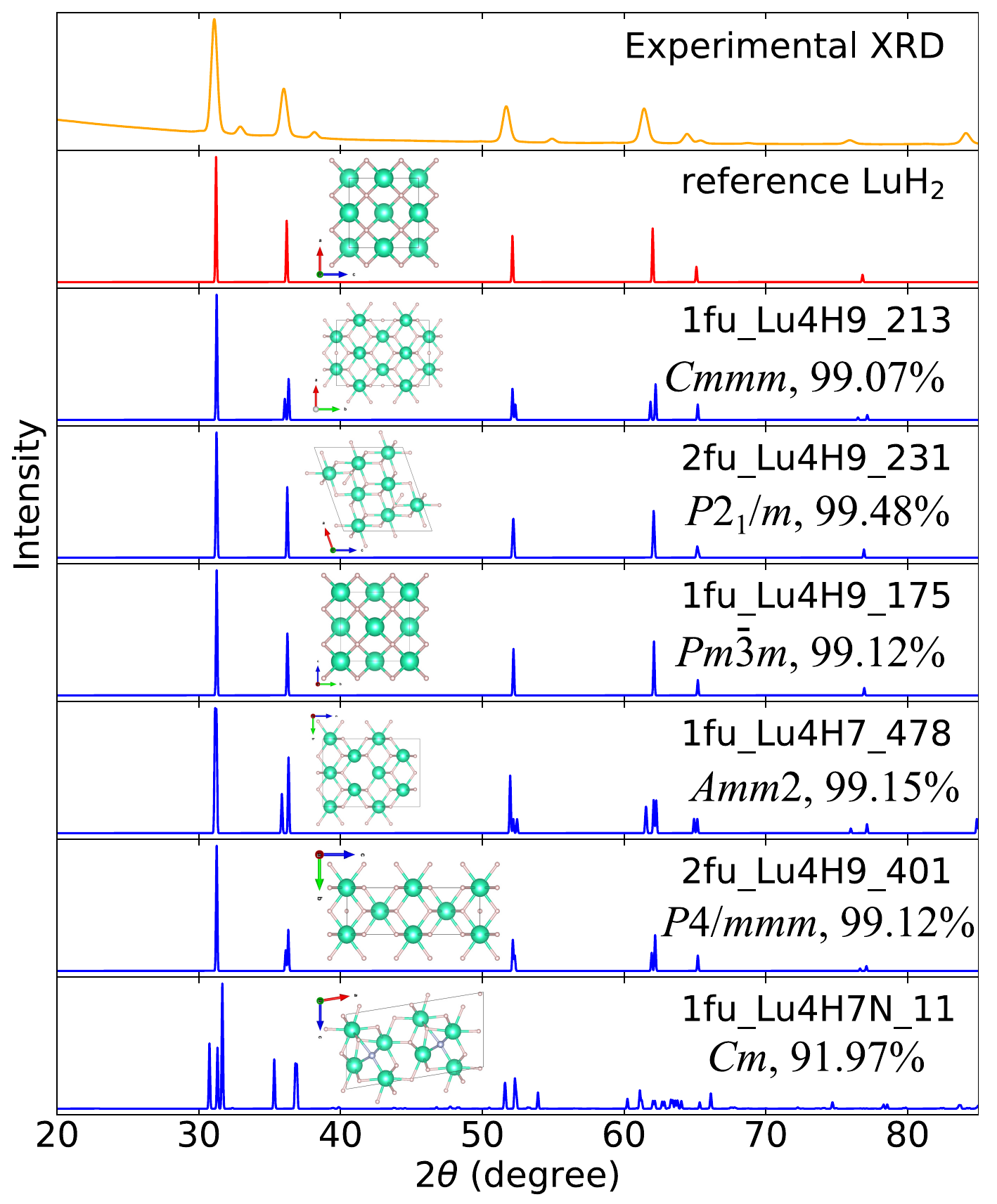}
\caption{\label{fig:XRD_comparison}  
\textbf{XRD analysis and crystal structures.} The simulated XRD of predicted structures are compared with the reference state ${Fm\bar{3}m}$ LuH$_2$ and the experimental XRD in Ref.~\cite{Dias2023Nature-LuNH}. Structure ID, space group, and XRD similarity with respect to ${Fm\bar{3}m}$ LuH$_2$ are displayed, along with crystal structure maps. 
}
\end{figure}

Among the binary lutetium hydrides that have been studied experimentally,  ${Fm\bar{3}m}$ LuH$_2$ has been suggested by Xie et al.~\cite{Xie2023-LuHN-MiaoLiu-ChinePhysLett} and Ming et al.~\cite{Ming_2023_Nature_HaihuWen} to have the most similar XRD pattern as the one measured by Dasenbrock-Gammon et al.~\cite{Dias2023Nature-LuNH}. 
We thus compare the XRD of LuH$_2$ with all 638 predicted phases from our high-throughput screening with ${H_{\rm hull}}$ $\le$ 0.24 eV${\cdot}$atom$^{-1}$ at 1 GPa.
In order to compare the XRDs of different structures quantitatively, the similarity between two XRDs is computed according to the correlation function implemented in PyXtal~\cite{pyxtal-2021QiangZhu}. For simpler comparison, the XRD of ${Fm\bar{3}m}$ LuH$_2$ is set as the reference and its simulated XRD is shown in Fig.~\ref{fig:XRD_comparison} as a comparison with the experimental XRD reported by Dasenbrock-Gammon et al.~\cite{Dias2023Nature-LuNH}.
The XRD similarity percentages ($\%$) for the 57 dynamically stable metallic structures are shown in Table.~\ref{table:57metallic-Tc-XRD}, while a comprehensive list of the 214 metallic states, irrespective of their dynamical stability, is available in Table~\ref{table:214metallic-Tc-XRD}.
By analyzing the 214 metallic phases, we find that 48 of them show strong XRD similarity percentages larger than 90$\%$. In contrast to the metallic phases, we do not find any structure among the 424 insulating phases showing XRD similarity percentages $\ge$ 90$\%$.
Out of these 48 metallic phases showing strong similarity in XRD, 24 of them are dynamically stable, including 9 Lu$_4$H$_7$ (mean XRD similarity $\sim$ 97.68$\%$), 14 Lu$_4$H$_9$ (mean XRD similarity $\sim$ 97.26$\%$), and 1 Lu$_4$H$_7$N$_{11}$ (XRD similarity $\sim$ 91.97$\%$). In Fig.~\ref{fig:XRD_comparison}, the dynamically stable binary hydrides with XRD similarity percentages larger than 99$\%$ are presented, together with the dynamically stable ${Cm}$ Lu$_4$H$_7$N (ID: 1fu\_Lu4H7N\_11) displaying XRD similarity higher than 90$\%$. In addition, the crystal structures are also shown in the inset of Fig.~\ref{fig:XRD_comparison}. Observing the space groups and the crystal structures in Fig.~\ref{fig:XRD_comparison}, although all six metallic phases show very high XRD similarity, the crystal structure varies from cubic lattice with high symmetry of ${Pm\bar{3}m}$ to the monoclinic lattice with low symmetry of ${Cm}$. The high XRD similarity can even occur in the triclinic structures. For instance, the 2fu\_Lu4H11N\_570 structure, included in Table~\ref{table:214metallic-Tc-XRD}, shows a large XRD similarity of 93.23$\%$ despite its space group is $P1$. 

The pronounced similarity in XRD can be attributed to the Lu sublattice, which does not differ significantly from the fcc one even if the symmetry reduction is considerable. In Figs.~\ref{fig:metallic-phonopy-stable-Lu4H7}, \ref{fig:metallic-phonopy-stable-Lu4H9} and \ref{fig:metallic-phonopy-stable-LuHN} which show the phonon spectra and XRD similarity of the 57 dynamically stable metallic phases, we find that 20 structures exhibiting high XRD similarity greater than 95$\%$ share a common feature that their distribution of phonons in frequencies resemble that of ${Fm\bar{3}m}$ LuH$_2$ at 1 GPa displayed in Fig.~\ref{fig:luh2-phonon-1GPa-phonopy}a. The specific characteristic is that most phonon branches are separated into two distinct frequency regions, in which some phonons are distributed below 6 THz, which have mainly a Lu character, and other branches are located around 32 THz. By comparing these spectra with the one of LuH$_2$ (see Fig.~\ref{fig:luh2-phonon-1GPa-phonopy}b), we find that all frequencies around 32 THz in these compounds are associated to vibrations of H atoms around tetrahedral interstitial sites. 
This means that, despite the notable variations in space groups among the 20 metallic states in Table~\ref{table:57metallic-Tc-XRD}, the high XRD similarity and shared characteristics in the phonon spectra imply that these structures are similar. 
Therefore, we have performed an in-depth analysis of the occupation of H atoms and the configuration of Lu sublattices of the 20 metallic structures.
Fig.~\ref{fig:H_sites_Lu_lattice} illustrates the number of H atoms that are situated at tetrahedral and octahedral sites in different configurations of Lu sublattice. In binary Lu$_4$H$_7$ compounds, all hydrogen atoms occupy tetrahedral sites as expected, and the Lu sublattice forms the fcc structure except for 1fu\_Lu4H7\_64, exhibiting a body-centered tetragonal sublattice displayed in Fig.~\ref{fig:Lu-sublattice}a). When it comes to Lu$_4$H$_9$, there are always eight H atoms located at tetrahedral sites and one H atom positioned at an octahedral site. The fcc Lu sublattice is also observed in these Lu$_4$H$_9$ compounds although 2fu\_Lu4H9\_225 has a defective fcc Lu sublattice (see Fig.~\ref{fig:Lu-sublattice}b)  that is significantly distorted compared to the fcc Lu sublattice in ${Fm\bar{3}m}$ LuH$_2$. Our analysis suggests that the majority of metallic stable phases with high XRD similarity are derived from the cubic LuH$_2$, in which interstitial tetrahedral hydrogen atoms can be widely observed in fcc Lu sublattices.

Although Dasenbrock-Gammon et al.~\cite{Dias2023Nature-LuNH} claimed their samples were likely composed of ${Fm\bar{3}m}$ and ${Immm}$ phases based on their XRD measurements, our analysis of XRD similarity and the arrangements of Lu sublattice suggest that XRD technique is not very suitable for identifying the crystal structure of lutetium hydrides. As evidenced above, we have identified tens of dynamically stable metallic phases that possess similar XRD features, even though some of them have a considerably lower symmetry.

\subsection{Potential hosts for high-temperature superconductivity}\label{subsec:epc} 

Failing to find any candidate with $H_{\rm hull}$ $<$ 0.24 eV$\cdot$atom$^{-1}$ that can host high-temperature superconductivity in the Lu-H-N system at 1 GPa, we study high-symmetry crystal structures with potential high electron-phonon interaction that may be metastable even if farther above the convex hull: ${Pm\bar{3}m}$ Lu$_4$H$_{11}$N, $Im\bar{3}m$ LuH$_6$, and $Fm\bar{3}m$ LuH$_{10}$. As shown in Fig.~\ref{fig:luhn_phase_diagram}b, ${Pm\bar{3}m}$ Lu$_4$H$_{11}$N is located  0.27 eV$\cdot$atom$^{-1}$ above the convex hull (structure ID: 1fu\_Lu4H11N\_137) at 1 GPa. 
This structure can  be derived from LuH$_3$ by generating the conventional cell of $Fm\bar{3}m$ LuH$_3$ and substituting one of the octahedral H atoms with N. 
In the harmonic approximation, this Lu$_4$H$_{11}$N, whose space group is ${Pm\bar{3}m}$, is unstable at 1 GPa.
The other two high-symmetry binary structures considered are LuH$_6$ and LuH$_{10}$, which are artificially constructed based on the already known high-temperature hydrogen-based superconductors $Im\bar{3}m$ CaH$_6$~\cite{PRL2022-CaH6-exp-YanmingMA} and $Fm\bar{3}m$ LaH$_{10}$~\cite{Errea2020-LaH10Nature}. 
It is noted that the crystal structure of $Im\bar{3}m$ LuH$_6$ has been theoretically reported in Ref.~\cite{LuH_Pickard_CPS}. At 1 GPa, the phase diagram in Fig.~\ref{fig:luhn_phase_diagram}c shows that $Im\bar{3}m$ LuH$_6$ and $Fm\bar{3}m$ LuH$_{10}$ are located around 0.64 and 0.77 eV$\cdot$atom$^{-1}$ above the convex hull, implying a highly unstable nature at 1 GPa and 0 K. They are also both dynamically unstable at this pressure at the harmonic level. The zero-point energy is not enough to make any of these structures energetically competitive at 1 GPa. 

In order to investigate the impact of quantum anharmonic effects on the dynamical stability of these high-symmetry structures, we relax them within the stochastic self-consistent harmonic approximation (SSCHA)~\cite{Errea-PRB2014-SSCHA, Monacelli-JPCM2021-SSCHA, Bianco-PRB2017-SSCHA,monacelli2018pressure} at 300 K at different pressures. This completes the prior study performed on the dynamical stability of the potential parent $Fm\bar{3}m$ LuH$_2$ and $Fm\bar{3}m$ LuH$_3$ phases~\cite{dangic2023-arxiv-color-change-parent-structure}. In order to assess the dynamical stability of these phases we calculate the phonons derived from the Hessian of the SSCHA free energy and check for the presence of imaginary phonon modes~\cite{Bianco-PRB2017-SSCHA}.

\begin{figure*}[h!]
    \centering
    \includegraphics[width=\linewidth]{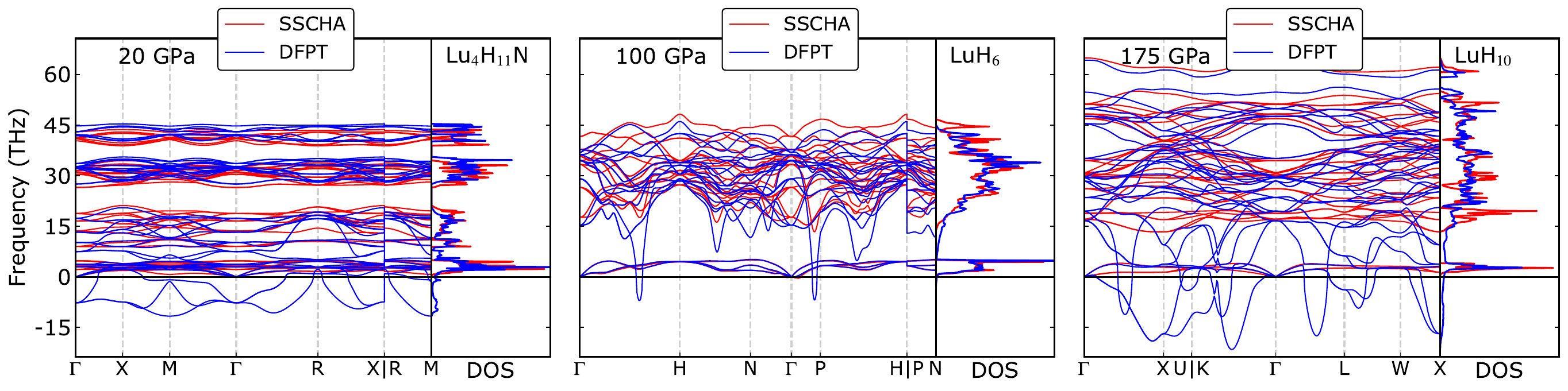}
    \caption{Harmonic and SSCHA free energy Hessian phonon spectra  of Lu$_4$H$_{11}$N, LuH$_6$, and LuH$_{10}$.}
    \label{fig:hessians}
\end{figure*}

 Our SSCHA analysis shows that ${Pm\bar{3}m}$ Lu$_4$H$_{11}$N becomes dynamically stable at around 20 GPa and 300 K (see Fig.~\ref{fig:hessians}), which is comparable to the stability range of LuH$_3$ (6 GPa and 300 K)~\cite{dangic2023-arxiv-color-change-parent-structure}. The phonon band structure shows five distinctive regions: Lu-dominated modes below 5 THz, N-dominated modes between 8 and 11 THz, and three regions of H-dominated modes above 11 THz. The highest phonon frequency of this system (43.7 THz) is slightly larger than in the case of LuH$_3$ (40 THz at 20 GPa and 300 K). Utilizing isotropic Migdal-Eliashberg equations, we estimate the superconducting critical temperature in this material to be 100 K at this pressure. This is a significant improvement compared to the potential parent compound LuH$_3$ at the same pressure, $T_c$ = 30 K. The element projected Eliashberg spectral functions in Fig.~\ref{fig:comp_a2f} shows that most of the electron-phonon coupling comes from H dominated modes, with very small contributions from Lu and N. The major effect of N doping in this system thus comes from breaking the symmetry of tetrahedral sites occupied by H, leading to a drift of H atoms away from these positions. This in turn has a large impact on the phonon frequencies and the value of the electronic density of states at the Fermi level in this system and consequently the electron-phonon coupling. 
 In Table~\ref{table:comparison-LuH3-Lu4H11N}, the crystal structure parameters and the Wyckoff positions of the atoms of ${Pm\bar{3}m}$ Lu$_4$H$_{11}$N are compared with those of $Fm\bar{3}m$ LuH$_{3}$. It is clear that the H atoms, which are originally located at the high-symmetry Wyckoff sites (0.25, 0.25, 0.25) in LuH$_3$ are shifted to lower symmetry Wyckoff sites (0.23835, 0.23835, 0.23835). In Fig.~\ref{fig:comparison-LuH3-Lu4H11N}, off-centering displacements of the hydrogen atoms at the tetrahedral sites are schematically shown to demonstrate the effect of N. Because the nitrogen atom replaces the hydrogen atom located at the octahedral site, the larger atomic radius pushes the hydrogen atoms at the tetrahedral sites outwards.

The binary compound $Im\bar{3}m$ LuH$_6$ has been predicted as a high-temperature superconductor at 100 GPa without the inclusion of ionic quantum anharmonic effects~\cite{LuH_Pickard_CPS}. We fully relax this compound at different pressures within the SSCHA to determine the phase diagram. Although we find it unstable at 100 GPa in the harmonic approximation, it is stabilized with anharmonic and quantum effects at 300 K. The instability in the harmonic approximation is localized at a singular $\mathbf{q}$-point, which in Ref.~\cite{LuH_Pickard_CPS} is also showing significant softening. The \textbf{q} point that shows significant softening in Ref.~\cite{LuH_Pickard_CPS} appears in an $8\times 8\times 8$ grid, which could explain why the instability is not present in the previous study that uses instead a $6\times 6\times 6$ grid. The calculated electron-phonon coupling constant is huge, which explains the softening of this phonon mode. 
We find nearly room-temperature superconductivity at 100 GPa with $T_c$ of 246 K in this structure, similar to the value reported in Ref.~\cite{LuH_Pickard_CPS}, where a $T_c$ of 273 K was predicted at 100 GPa at the harmonic level.

The Fm$\bar{3}$m LuH$_{10}$ structure is the one adopted by the high-temperature superconductor LaH$_{10}$. We find that the free energy Hessian does not show imaginary frequencies above 175 GPa and 300 K indicating a metastable state. We calculate the electron-phonon coupling for this structure and found that the onset of superconductivity happens practically at room temperature, $T_c$ = 289 K ($\sim$ 16 $^{\circ}$C). In comparison to another polymorph of LuH$_{10}$ with a space group ${P63/mmc}$, which can also be stabilized at 200 GPa in the harmonic approximation and has a theoretically reported $T_c$ of 134-152 K~\cite{PhysRevLett.125.217001-LuH10}, $Fm\bar{3}m$ LuH$_{10}$ phase in our study not only refreshes the record of highest $T_c$ of LuH$_{10}$ but it can also be stabilized dynamically at a reduced pressure. 

In order to confirm the capacity of the networking value model to predict $T_c$s used in the high-throughput calculations, we  also estimate $T_c$ for these high-symmetry structures with it for the SSCHA structures. We found that the $T_c$ of ${Fm\bar{3}m}$ LuH$_3$ and ${Pm\bar{3}m}$ Lu$_4$H$_{11}$N at 20 GPa are 47.46 and 99.67 K, respectively. Furthermore, the networking value model predicts that LuH$_6$ at 100 GPa and LuH$_{10}$ at 175 GPa are superconductors with a $T_c$ of 296.85 K and 389.50 K, respectively. Remarkably, the $T_c$ values are very close to those obtained with {\it ab initio} electron-phonon coupling calculations, especially for LuH$_3$ and  Lu$_4$H$_{11}$N. This implies that the networking value model can be applicable to the estimation of $T_c$ of superhydrides, even for those that have never been reported. Consequently, it justifies the use of the networking value model as a rapid estimator of the $T_c$ of the structures obtained in the high-throughput screening (i.e. the results in Table~\ref{table:57metallic-Tc-XRD}).

In order to estimate whether these high-symmetry structures may be identified by diffraction experiments, we compute the XRD similarity of ${Pm\bar{3}m}$ Lu$_4$H$_{11}$N, $Im\bar{3}m$ LuH$_6$, and $Fm\bar{3}m$ LuH$_{10}$ at 1 GPa with reference to the cubic LuH$_2$. The results show that while $Im\bar{3}m$ LuH$_6$ and $Fm\bar{3}m$ LuH$_{10}$ are very different, with a similarity of 35$\%$ and 58$\%$, respectively, ${Pm\bar{3}m}$ Lu$_4$H$_{11}$N has a similarity of 95$\%$, which remarks that it may be indistinguishable in diffraction experiments from $Fm\bar{3}m$ LuH$_{2}$ and may be consistent with the observed XRD pattern in Ref.~\cite{Dias2023Nature-LuNH}.

\begin{figure*}[h!]
    \centering
    \includegraphics[width=\linewidth]{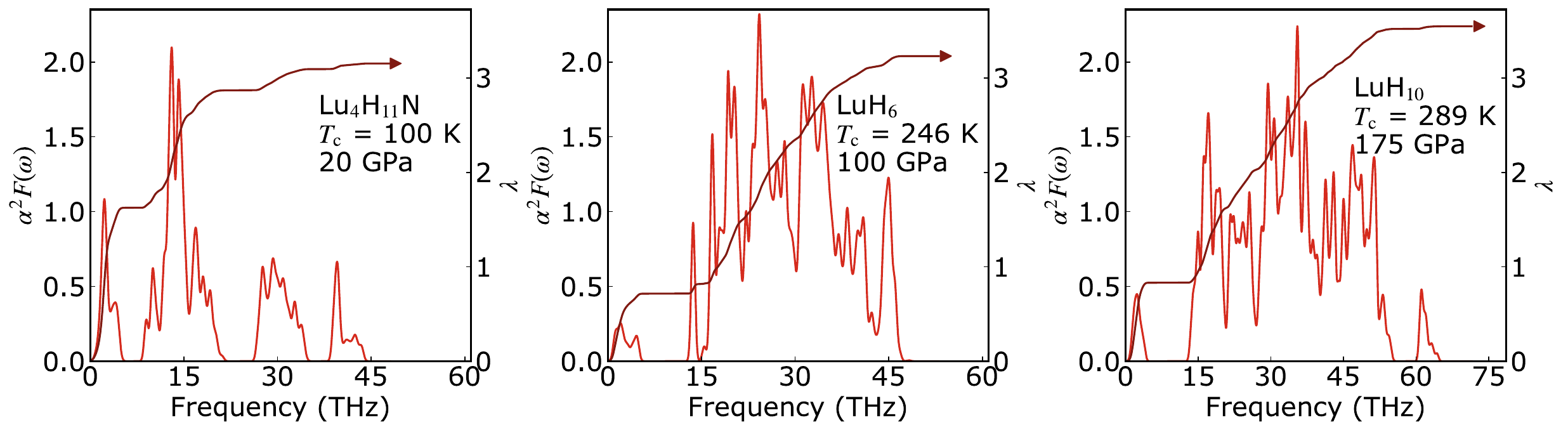}
    \caption{Isotropic Eliashberg spectral function ${\alpha^2F(\omega)}$ and integrated electron-phonon coupling constant $\lambda$ in Lu$_4$H$_{11}$N, LuH$_6$ and LuH$_{10}$.}
    \label{fig:a2Fs}
\end{figure*}

\section{Conclusions}

In conclusion, we have performed a comprehensive study by combining a high-throughput structure screening, a rapid estimator of $T_c$, and electron-phonon coupling calculations including quantum anharmonic effects to explore the feasibility of near-ambient superconductivity in the Lu-N-H systems. Our study suggests that the presence of nitrogen favors the formation of insulating states rather than metallic ones, and destabilizes the lutetium hydride systems by significantly shifting them away from the convex hull. As a result, the majority of identified dynamically stable metallic phases in our investigation are binary hydrides, which are not far from the parent $Fm\bar{3}m$ LuH$_{2}$, rather than ternary hydrides. Furthermore, we did not observe high-temperature superconductivity in all of the studied structures at 1 GPa within a reasonable threshold for metastability. We, therefore, propose that in order to have metallic and superconducting states in lutetium hydrides it is better to avoid nitrogen doping. 
Despite the absence of near-ambient superconductivity, the combined effects of pressure and quantum anharmonicity kindles the hope of high-temperature superconductivity in the Lu-N-H systems by realizing a $T_c$ of 100 K at only 20 GPa in cubic Lu$_4$NH$_{11}$. This structure is similar to $Fm\bar{3}m$ LuH$_{3}$, but with one out of four H atoms in octahedral sites substituted by a nitrogen atom. Even if this structure is far from being thermodynamically stable at 1 GPa and is dynamically unstable below 20 GPa in our calculations, it is practically indistinguishable from the parent $Fm\bar{3}m$ LuH$_{2}$ compound in diffraction experiments. Thus, this structure or variations of it provide, if any, the only possible high-$T_c$ structure at low pressures in the Lu-H-N system. At higher pressures, above 100 GPa, CaH$_6$-like LuH$_6$ and LaH$_{10}$-like LuH$_{10}$ are superconductors with critical temperatures around room temperature, but with an XRD pattern incompatible with experiments.

\section{Methods}\label{sec:DFT-methods}

\subsection{High-throughput crystal structure prediction}

State-of-the-art crystal structure prediction methods, i.e., the evolutionary algorithm implemented in CrySPY~\cite{Yamashita2021CrySPY} and the particle swarm algorithm implemented in CALYPSO~\cite{Wang2010,Wang2012}, were combined to predict crystal structures. 
Crystal structure predictions were performed within first-principles density functional theory (DFT) calculations using the Vienna Ab initio Simulation Package
(VASP)~\cite{Kresse-PRB-1996,Kresse1996}. The generalized gradient approximation within the parametrization of Perdew, Burke, and Ernzerhof~\cite{PBE-Perdew1996} was used with a Hubbard $U$ correction in the Dudarev’s form~\cite{Dudarev-LDAU-PRB1998}
to improve the accuracy of the energies of the Lu $f$ states. An acceptable value of $U$ = 5.5 eV, commonly used to account for the localized $f$-states of the lanthanide systems~\cite{Dias2023Nature-LuNH,PhysRevB.94.014104}, was used.
The plane wave energy cutoff was set to 450 eV during crystal structure predictions.
The \textbf{k}-point grid is generated based on the specific structure by Pymatgen~\cite{Jain2011_pymatgen} with a relatively high grid density of 60 points per \AA$^{-3}$ of reciprocal cell volume.
We have screened over 15,000 crystal structures for the Lu-N-H system.
To address the electronic properties of the structures within  0.1 eV per atom above the convex hull, the energy cutoff was improved to 550 eV with an improved \textbf{k}-point grid density of 80 points per \AA$^{-3}$.

To calculate the hydrogen fraction of the total DOS ${H_{\rm DOS}}$ at the Fermi level, we used Sumo~\cite{Ganose2018-sumocode} to extract the hydrogen DOS and the total DOS. The TcESTIME code has been used to estimate the superconducting $T_c$ based on the networking value model~\cite{Belli2021-network-value-NC}. The XRD simulation and comparison of the XRD similarity were performed by PyXtal~\cite{pyxtal-2021QiangZhu}.

In the phonon calculations for the crystal structures predicted from high-throughput structure screening, the VASP DFT calculations were combined with the supercell and finite displacement methods implemented in Phonopy~\cite{Togo-phonopy2015}. 
Initially, the unit cells from the crystal structure predictions were further optimized with a cutoff energy of 550 eV and a \textbf{k}-point grid density of 130 per \AA$^{-3}$ of reciprocal cell volume until the energy convergence reaches 10${^{-8}}$ eV and forces of each atom were less than 10${^{-3}}$ eV${\cdot}$\AA$^{-1}$. Subsequently, the optimized cells were expanded to supercells for force calculations in DFT. 
However, considering there were many structures to be examined, it has become infeasible to consider particularly large supercells. It is noted that atomic interactions in most cases were significantly decreased with increased interatomic distances, and thus a cutoff distance of 7.2 \AA~was considered to be proper in setting up the supercells. Specifically, if the lattice parameter ($a$, $b$, or $c$) of the unit cell was smaller than 7.2 \AA, it was expanded twice, otherwise, it remained unchanged. With this constraint, the  214 metallic phases with ${H_{\rm hull}}$ $\leq$ 0.24 eV${\cdot}$atom$^{-1}$ from the high-throughput crystal structure predictions resulted in more than 165,000 supercells for DFT calculations.

\subsection{Electron-phonon coupling calculations}\label{subsec:epc}

We relaxed the $Pm\bar{3}m$ Lu$_4$H$_{11}$N, $Im\bar{3}m$ LuH$_6$ and $Fm\bar{3}m$ LuH$_{10}$ using the stochastic self-consistent harmonic approximation method~\cite{Errea-PRB2014-SSCHA, Monacelli-JPCM2021-SSCHA, Bianco-PRB2017-SSCHA, monacelli2018pressure} on $2\times 2\times 2$ supercells. The number of configurations used in the minimization of the free energy was 400 for Lu$_4$H$_{11}$N, and 200 for LuH$_6$ and LuH$_{10}$. The calculation for the free energy Hessian phonons needed to confirm the dynamical stability of final structures was performed with 5000 configurations for each structure. 

To calculate superconducting critical temperature for these compounds we performed electron-phonon calculations for the structures obtained through the SSCHA minimization of total free energy using density functional perturbation theory (DFPT) method as implemented in Quantum Espresso~\cite{QE-2009,QE-2017}. Electron-phonon coupling constants were calculated on a $4\times 4\times 4$ $\mathbf{q}$ point grid for Lu$_4$H$_{11}$N, and an $8\times 8\times 8$ grid for LuH$_6$ and LuH$_{10}$. The average of the electron-phonon matrix elements over the Fermi surface was done on $24\times 24\times 24$ $\mathbf{k}$ point grid and 0.012 Ry smearing for Lu$_4$H$_{11}$N and $42\times 42\times 42$ $\mathbf{k}$ point grid and 0.008 Ry smearing for LuH$_6$ and LuH$_{10}$. Unfortunately, large electron-phonon coupling makes the full convergence of results in the LuH$_6$ compound very challenging. However, the estimation of critical temperature is quite robust and does not change more than 20 K between the two highest \textbf{k}-point grids ($36^3$ and $42^3$) and two lowest smearing values (0.008 and 0.012 Ry).
The Eliashberg spectral function was calculated using phonon frequencies obtained from free energy Hessian. The solution of the isotropic Migal-Eliashberg equation was obtained with the cutoff for Matsubara frequencies of 10 times the highest phonon frequency and the reduced Coulomb interaction of $\mu^* = 0.16$. The calculation of the electron-phonon coupling for $Fm\bar{3}m$ LuH$_3$ was done following the parameters previously used in Ref.~\cite{dangic2023-arxiv-color-change-parent-structure}.

\section{Code availability}

The high-throughput crystal structural predictions were carried out
using the proprietary code VASP~\cite{Kresse1996,Kresse-PRB-1996}, with the
combination of CALYPSO~\cite{Wang2010, Wang2012} and
CrySPY~\cite{Yamashita-CrySPY}.  CALYPSO (http://www.calypso.cn/) is
freely distributed for academic users under the license of Copyright
Protection Center of China (registration No. 2010SR028200 and
classification No. 61000-7500).  CrySPY
(https://github.com/Tomoki-YAMASHITA/CrySPY) is released under the
Massachusetts Institute of Technology (MIT) License and is open
source.  The phonon and electron-phonon properties were calculated by
Phonopy~\cite{Togo-phonopy2015}, Quantum ESPRESSO~\cite{QE-2009,QE-2017} and SSCHA~\cite{Errea-PRB2014-SSCHA, Monacelli-JPCM2021-SSCHA, Bianco-PRB2017-SSCHA,monacelli2018pressure}.  Phonopy
(https://github.com/atztogo/phonopy) is openly released under the
BSD-3-Clause License.  
The SSCHA code (https://github.com/SSCHAcode/python-sscha) is open source and is based on the GNU General Public License v3.0.
Quantum ESPRESSO (https://www.quantum-espresso.org) is an open source suite of computational tools with GNU General Public License v2.0. 
The crystal structure visualization software
VESTA~\cite{momma2011vesta} is distributed free of charge for academic
users under the VESTA License (https://jp-minerals.org/vesta/en/). The visualization tools for phonon and electronic properties, Sumo (https://github.com/SMTG-UCL/sumo) and Phtools (https://github.com/yw-fang/phtools), are both based on the MIT License.
The TcESTIME code is freely available at https://www.lct.jussieu.fr/pagesperso/contrera/tcestime/index.html. The PyXtal code is freely available under the MIT License (https://github.com/qzhu2017/PyXtal).

\section{Data availability}
The data that support all the findings of this study are available in the manuscript and in the Supplementary Materials. In addition, some raw data are provided at Zenodo open data repository (reserved DOI: 10.5281/zenodo.8140540). This repository is accessible upon request during peer review of this manuscript in any journal. Once the manuscript is accepted, this dataset is freely open to all.
This repository contains:
\begin{enumerate}
    \item A complete list of 214 metallic phases regardless of the dynamical stability (output-including-XRD-regardless-stability-240meV.xlsx) recording the information of ID, $T_c$, networking value, H$_{DOS}$, XRD similarity, space group, etc.
    \item Database of all the structures in VASP format whose enthalpy distance to the convex hull is less or equal to 0.24 eV per atom.
    \item Both structures and dynamical matrix files in Quantum Espresso format for the electron-phonon coupling calculations for cubic Lu$_4$H$_{11}$N, ${Im\bar{3}m}$ LuH$_6$, and ${Pm\bar{3}m}$ LuH$_{10}$. 
\end{enumerate}

\section{Acknowledgments}

This project is funded by the European Research Council (ERC) under the European Union's Horizon 2020 research and innovation program (Grant Agreement No. 802533) and the Department
of Education, Universities and Research of the Eusko
Jaurlaritza and the University of the Basque Country
UPV/EHU (Grant No. IT1527-22).
We acknowledge PRACE for awarding us access to the EuroHPC supercomputer LUMI located in CSC's data center in Kajaani, Finland through EuroHPC Joint Undertaking (EHPC-REG-2022R03-090).

\section{Author contributions}
Y.-W.F. and I.E. conceived the study and planned the research.
Y.-W.F. and {\DJ}.D. performed the theoretical calculations.
Y.-W.F. wrote the manuscript with substantial input from {\DJ}.D. and I.E.
I.E. secured research grants and computational resources.

\section{Competing interests}
The authors declare no competing interests.

\section{Additional information}
Materials \& Correspondence should be addressed to Y.-W.F.

\clearpage
\newpage
\appendix

\setcounter{figure}{0} 
\renewcommand\thefigure{S\arabic{figure}} 
\renewcommand{\theHfigure}{S\arabic{figure}}

\setcounter{table}{0} 
\renewcommand\thetable{S\arabic{table}} 

\section{Properties of lowest-enthalpy ternary state at 1 GPa: ${P2_1/m}$ LuH$_2$N }
\begin{figure}[ht]
\includegraphics[angle=0,width=0.99\textwidth]{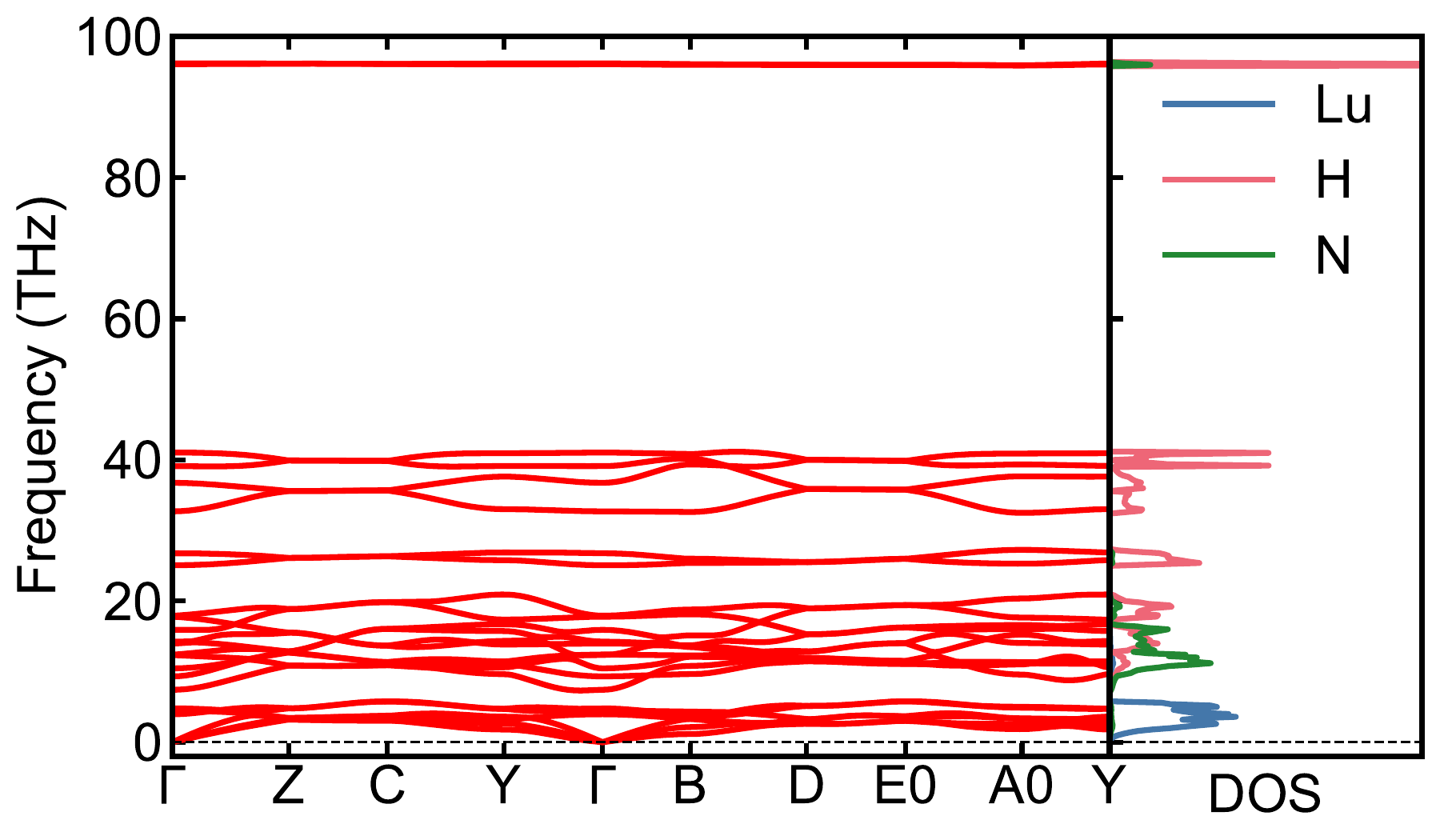}
\caption{\label{fig:2fu_LuH2N_389_phonon_supercell222}  
\textbf{The phonon dispersion and element projected phonon DOS of ${P2_1/m}$ LuH$_2$N. } 
}
\end{figure}

\begin{figure}[ht]
\includegraphics[angle=0,width=0.99\textwidth]{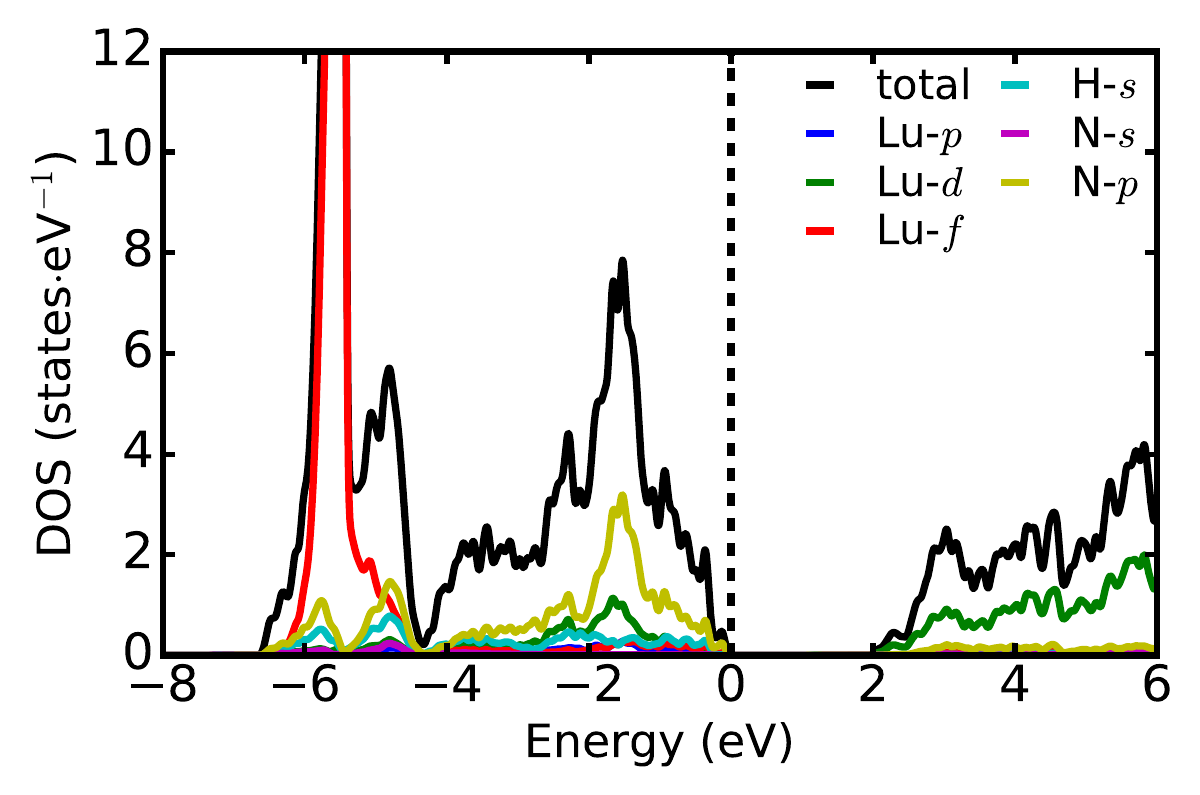}
\caption{\label{fig:2fu_LuH2N_389_e_dos}  
\textbf{The electronic density of states (DOS) of ${P2_1/m}$ LuH$_2$N (ID: 2fu\_LuH2N\_389).}  The major components of the DOS are shown with projections to orbitals. 
} 
\end{figure}

Fig.~\ref{fig:2fu_LuH2N_389_phonon_supercell222} shows the phonon dispersion and element projected phonon DOS of ${P2_1/m}$ LuH$_2$N (ID: 2fu\_LuH2N\_389) calculated by VASP and Phonopy. Fig.~\ref{fig:2fu_LuH2N_389_e_dos} displays the calculated electronic DOS.

\clearpage

\section{Phonon spectra of 57 stable metallic Lu-H and Lu-N-H phases in the harmonic approximation}
\begin{figure}[ht]
\includegraphics[angle=0,width=0.95\textwidth]{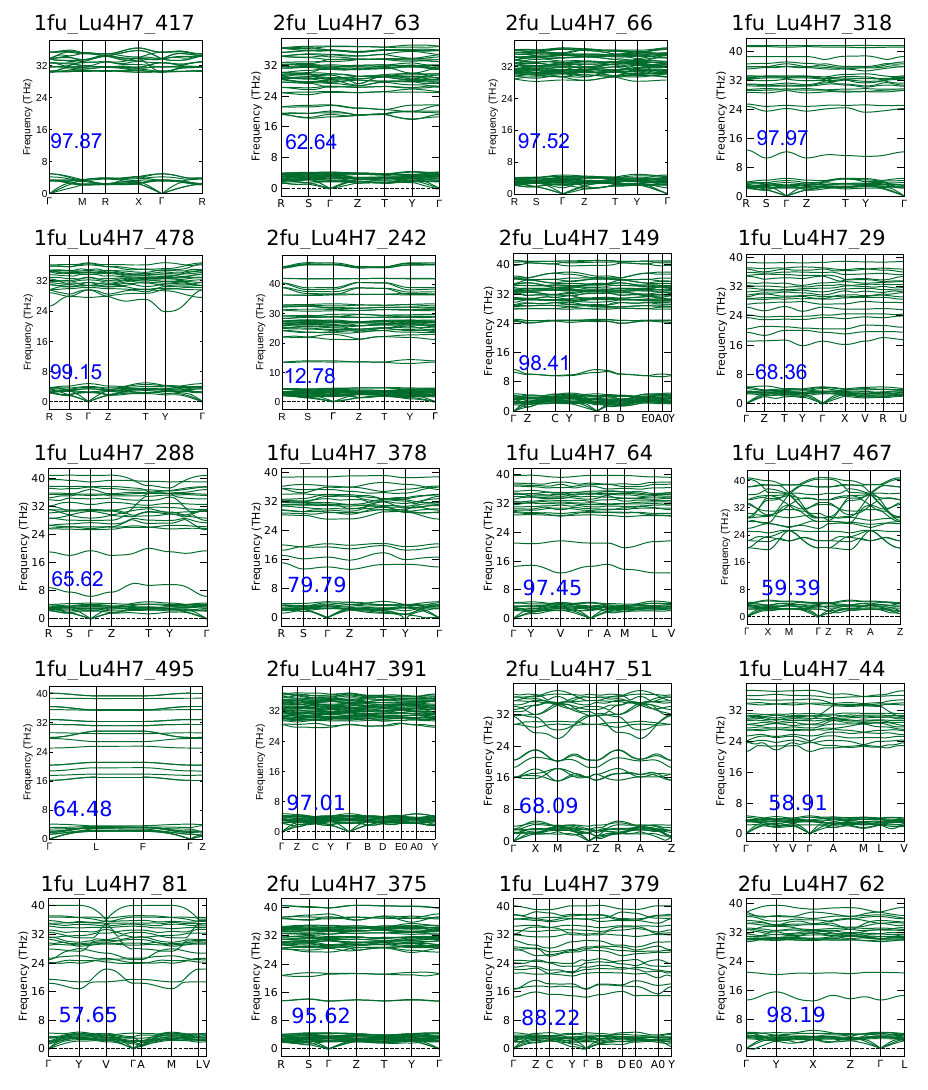}
\caption{\label{fig:metallic-phonopy-stable-Lu4H7}  
\textbf{Harmonic phonon spectra of 20 Lu$_4$H$_7$ compounds that are predicted to be dynamically stable by the supercell method.} The blue text shows the XRD similarity ($\%$). 
}
\end{figure}

\begin{figure}[t]
\includegraphics[angle=0,width=0.95\textwidth]{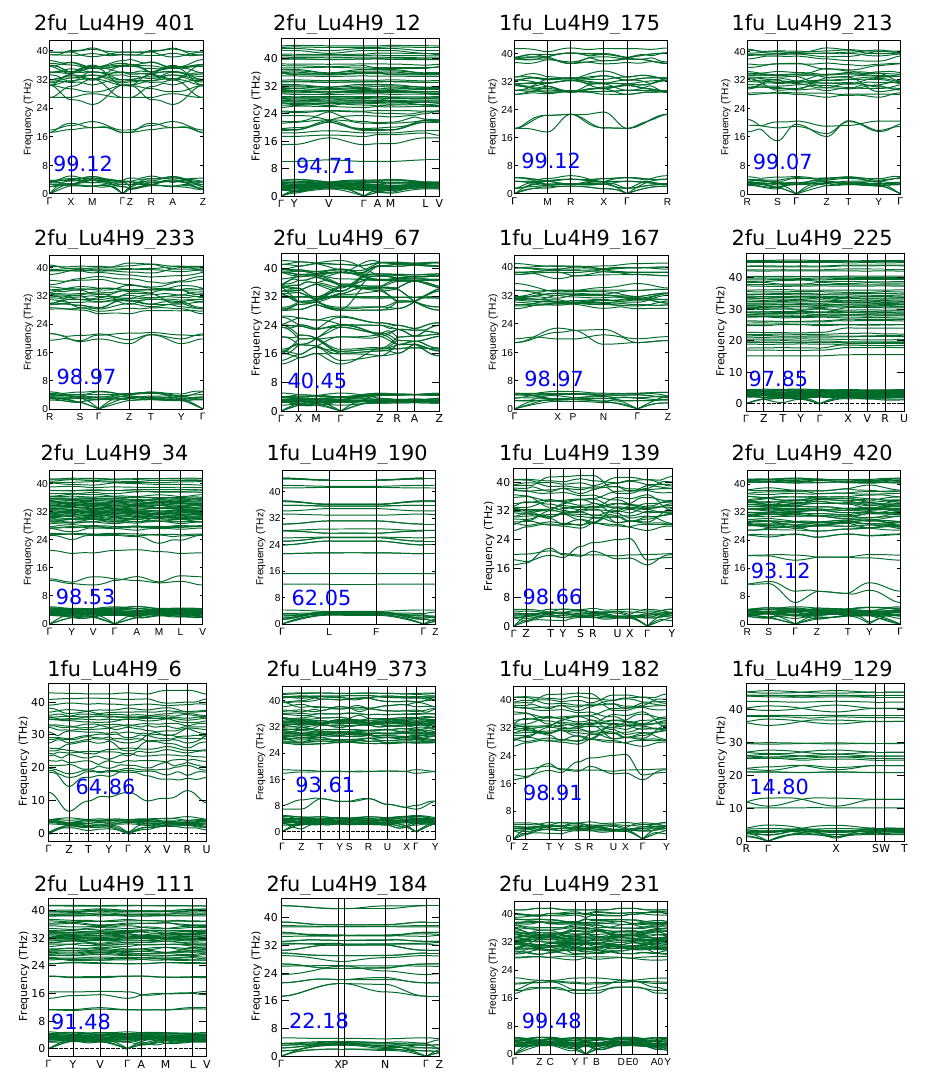}
\caption{\label{fig:metallic-phonopy-stable-Lu4H9}  
\textbf{Harmonic phonon spectra of 19 Lu$_4$H$_9$ compounds that are predicted to be dynamically stable by the supercell method.} The blue text shows the XRD similarity ($\%$).
}
\end{figure}

\begin{figure}[ht]
\includegraphics[angle=0,width=0.95\textwidth]{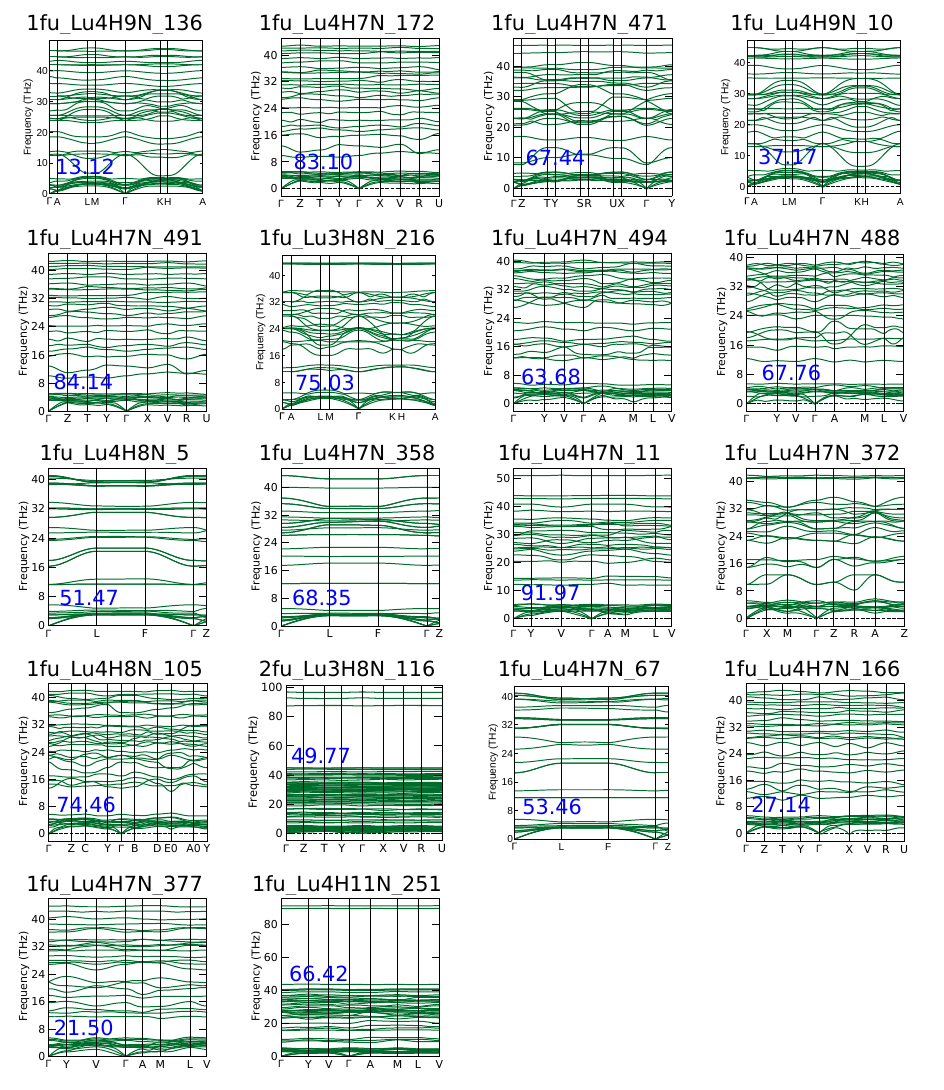}
\caption{\label{fig:metallic-phonopy-stable-LuHN}  
\textbf{Harmonic phonon spectra of 18 ternary Lu-N-H compounds that are predicted to be dynamically stable by the supercell method.}
The 18 ternary Lu-N-H compounds include 11 phases of Lu$_4$H$_7$N, 2 phase of Lu$_4$H$_9$N, 2 phases of Lu$_4$H$_8$N, 2 phases of Lu$_3$H$_8$N, and 
1 phase of Lu$_4$H$_{11}$N. The blue text shows the XRD similarity ($\%$).
}
\end{figure}

\subsection{Phonon spectra of ${Fm\bar{3}m}$ LuH$_2$ at 1 GPa}
\begin{figure}[ht]
\includegraphics[angle=0,width=0.95\textwidth]{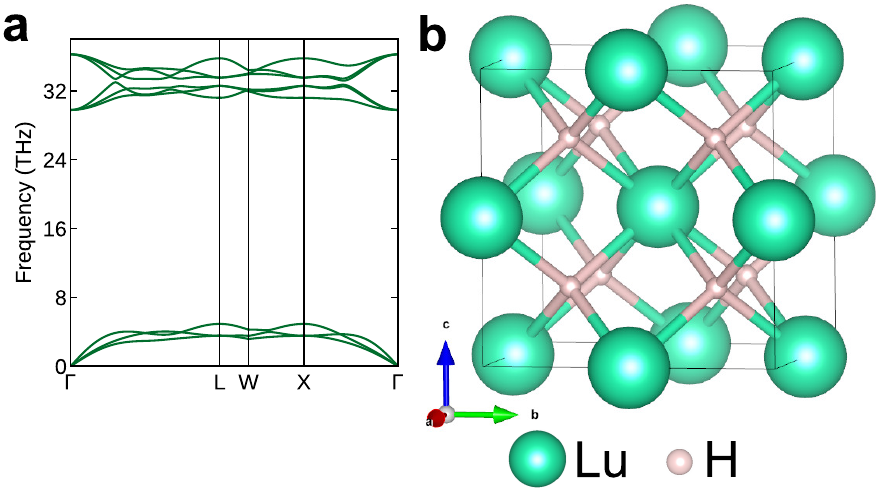}
\caption{\label{fig:luh2-phonon-1GPa-phonopy}  
\textbf{The properties of ${Fm\bar{3}m}$ LuH$_2$ at 1 GPa.
} \textbf{a} Harmonic phonon spectrum calculated with the finite displacements method. \textbf{b} Crystal structure at 1 GPa which shows all H atoms are located at the interstitial tetrahedral sites.
}
\end{figure}

Fig.~\ref{fig:luh2-phonon-1GPa-phonopy} shows the phonon spectra of ${Fm\bar{3}m}$ LuH$_2$ calculated by finite displacement method in Phonopy.

\clearpage
\newpage

\section{Networking-value-predicted $T_c$ of all the metallic phases within 0.24 eV$\cdot$atom$^{-1}$ above the convex hull regardless of the dynamical stability.}

\begin{center}
\captionsetup{width=\linewidth}
\begin{longtable}{@{\extracolsep{\fill}}ccccccc}
\caption{\label{table:214metallic-Tc-XRD}
All the metallic states at 1 GPa within 0.24 eV${\cdot}$atom$^{-1}$ above the convex hull are considered. The superconducting transition temperatures ($T_c$) are estimated with the networking value model in Ref.~\cite{Belli2021-network-value-NC}. ${H_{\rm hull}}$ represents the enthalpy distance above the convex hull. ID refers to the unique identifier of a structure entry in our database.  ${H_{\rm DOS}}$ is the hydrogen fraction of the total density of states at the Fermi level, $\phi$ refers to the networking value. XRD similarity refers to the degree of similarity in the simulated X-ray diffraction (XRD) patterns between the structure being analyzed and the reference structure (i.e. cubic LuH$_2$). In several cases where data is not available, \textit{N/A} is used.} 
\\
\hline
ID & $H_{\rm hull}$ (eV$\cdot$atom$^{-1}$) & Space group & ${H_{\rm DOS}}$ & $\phi$ & $T_c$ ($\pm$60 K)  & XRD similarity ($\%$) \\
\hline
\endfirsthead
\hline
ID & $H_{\rm hull}$ (eV$\cdot$atom$^{-1}$) & Space group & ${H_{\rm DOS}}$ & $\phi$ & $T_c$ ($\pm$60 K)  & XRD similarity ($\%$) \\
\hline
\endhead
\hline
\multicolumn{7}{r}{\textit{Continued on next page}} \\
\endfoot
\hline
\endlastfoot

1fu\_Lu3H8N\_216 & 0.24 & $P\bar{3}m1$ & 0.26 & 0.31 & 13.94 & 75.03 \\
1fu\_Lu4H9\_138 & 0.21 & $P\bar{4}3m$ & 0.05 & 0.44 & -2.68 & 88.97 \\
1fu\_Lu4H9\_204 & 0.21 & $Cm$ & 0.05 & 0.42 & -4.52 & 86.67 \\
1fu\_Lu4H11N\_462 & 0.21 & $Cm$ & 0.40 & 0.21 & -4.95 & 52.28 \\
1fu\_Lu4H9\_63 & 0.13 & $Cmm2$ & 0.03 & 0.47 & -8.83 & 85.62 \\
1fu\_Lu4H11N\_34 & 0.20 & $Pm$ & 0.38 & 0.20 & -10.6 & 36.88 \\
2fu\_Lu4H11N\_570 & 0.14 & $P1$ & 0.18 & 0.25 & -11.59 & 93.23 \\
1fu\_Lu4H11N\_156 & 0.24 & $Pm$ & 0.34 & 0.20 & -12.94 & 43.05 \\
4fu\_Lu2H5N\_142 & 0.20 & $Imma$ & 0.32 & 0.22 & -14.74 & 43.34 \\
2fu\_Lu4H9\_155 & 0.20 & $P1$ & 0.03 & 0.42 & -15.73 & 41.71 \\
1fu\_Lu3H8N\_419 & 0.22 & $Amm2$ & 0.31 & 0.20 & -17.64 & 76.31 \\
1fu\_Lu4H7\_495 & 0.17 & $R3m$ & 0.04 & 0.42 & -17.88 & 64.48 \\
2fu\_Lu4H7\_242 & 0.17 & $Cmc2_1$ & 0.04 & 0.40 & -19.35 & 12.78 \\
2fu\_Lu4H9\_373 & 0.05 & $Pba2$ & 0.03 & 0.39 & -19.56 & 93.61 \\
1fu\_Lu4H9N\_150 & 0.22 & $Cm$ & 0.04 & 0.37 & -22.13 & 80.64 \\
2fu\_Lu4H7\_205 & 0.23 & $P\bar{1}$ & 0.02 & 0.45 & -23.48 & 68.34 \\
1fu\_Lu4H9\_22 & 0.01 & $C2/m$ & 0.04 & 0.33 & -25.1 & 99.25 \\
2fu\_Lu4H9\_34 & 0.02 & $Cm$ & 0.04 & 0.33 & -25.32 & 98.53 \\
1fu\_Lu4H9\_213 & 0.01 & $Cmmm$ & 0.04 & 0.34 & -25.58 & 99.07 \\
2fu\_Lu4H9\_231 & 0.01 & $P2_1/m$ & 0.04 & 0.34 & -26.0 & 99.48 \\
1fu\_Lu4H9\_99 & 0.16 & $Cmm2$ & 0.04 & 0.34 & -26.2 & 63.02 \\
1fu\_Lu4H9\_86 & 0.05 & $C2/m$ & 0.04 & 0.32 & -26.89 & 99.26 \\
1fu\_Lu4H9\_114 & 0.12 & $P\bar{1}$ & 0.03 & 0.35 & -27.51 & 84.71 \\
1fu\_Lu4H9\_190 & 0.11 & $R3m$ & 0.05 & 0.30 & -28.28 & 62.05 \\
1fu\_Lu4H7N\_359 & 0.23 & $Fmm2$ & 0.04 & 0.37 & -28.4 & 69.44 \\
2fu\_Lu4H7\_467 & 0.17 & $C2/m$ & 0.02 & 0.42 & -29.02 & 99.45 \\
1fu\_Lu4H7\_312 & 0.18 & $Cm$ & 0.02 & 0.41 & -29.28 & 75.27 \\
2fu\_Lu4H7\_98 & 0.21 & $C2$ & 0.03 & 0.37 & -29.47 & 75.79 \\
1fu\_Lu4H7\_415 & 0.19 & $P4mm$ & 0.02 & 0.42 & -29.63 & 69.74 \\
1fu\_Lu4H9N\_202 & 0.14 & $P2/m$ & 0.26 & 0.18 & -29.73 & 23.46 \\
2fu\_Lu4H7\_340 & 0.22 & $Pbam$ & 0.03 & 0.39 & -30.11 & 40.43 \\
2fu\_Lu4H11N\_574 & 0.15 & $P1$ & 0.04 & 0.31 & -30.63 & 72.06 \\
2fu\_Lu4H9\_12 & 0.15 & $Cm$ & 0.03 & 0.32 & -31.37 & 94.71 \\
1fu\_Lu4H7\_166 & 0.20 & $Amm2$ & 0.02 & 0.45 & -31.62 & 97.01 \\
2fu\_Lu4H7\_223 & 0.20 & $Cc$ & 0.03 & 0.36 & -31.65 & 13.44 \\
2fu\_Lu4H9\_64 & 0.15 & $P2_1$ & 0.03 & 0.33 & -31.81 & 61.22 \\
1fu\_Lu4H7\_373 & 0.21 & $C2/m$ & 0.03 & 0.37 & -31.91 & 56.22 \\
2fu\_Lu4H7\_51 & 0.19 & $P4/mmm$ & 0.02 & 0.43 & -32.23 & 68.09 \\
1fu\_Lu4H9\_208 & 0.01 & $Pm\bar{3}m$ & 0.03 & 0.32 & -32.29 & 99.75 \\
2fu\_Lu4H9\_233 & 0.01 & $Cmmm$ & 0.04 & 0.30 & -32.3 & 98.97 \\
2fu\_Lu4H9\_190 & 0.01 & $Pmmm$ & 0.03 & 0.32 & -32.51 & 98.92 \\
2fu\_Lu4H9\_252 & 0.01 & $P\bar{1}$ & 0.04 & 0.30 & -32.61 & 99.16 \\
1fu\_Lu4H9\_139 & 0.06 & $Pmmm$ & 0.03 & 0.32 & -32.68 & 98.66 \\
2fu\_Lu4H7\_238 & 0.22 & $Cm$ & 0.03 & 0.36 & -32.7 & 28.71 \\
1fu\_Lu4H9\_175 & 0.01 & $Pm\bar{3}m$ & 0.03 & 0.32 & -32.85 & 99.12 \\
2fu\_Lu3H8N\_351 & 0.21 & $C2/m$ & 0.28 & 0.16 & -32.92 & 51.18 \\
2fu\_Lu4H9\_23 & 0.16 & $Cmmm$ & 0.03 & 0.34 & -32.97 & 75.46 \\
2fu\_Lu4H7\_391 & 0.04 & $Pc$ & 0.03 & 0.35 & -33.55 & 97.01 \\
2fu\_Lu4H7\_62 & 0.08 & $Fmm2$ & 0.03 & 0.35 & -33.57 & 98.19 \\
1fu\_Lu4H9\_215 & 0.01 & $Pmmm$ & 0.03 & 0.31 & -33.72 & 98.52 \\
1fu\_Lu4H7\_204 & 0.06 & $R\bar{3}m$ & 0.02 & 0.39 & -33.8 & 96.55 \\
1fu\_Lu4H9\_182 & 0.01 & $Pmmm$ & 0.03 & 0.31 & -34.12 & 98.91 \\
2fu\_Lu4H9\_221 & 0.11 & $Pc$ & 0.03 & 0.30 & -34.26 & 61.50 \\
2fu\_Lu4H7\_202 & 0.21 & $Cmc2_1$ & 0.03 & 0.33 & -34.3 & 43.18 \\
2fu\_Lu4H7\_149 & 0.09 & $P2_1/m$ & 0.03 & 0.33 & -34.5 & 98.41 \\
2fu\_Lu4H7\_274 & 0.18 & $Pmmm$ & 0.03 & 0.36 & -34.68 & 97.59 \\
2fu\_Lu4H7\_66 & 0.03 & $Cmcm$ & 0.03 & 0.34 & -34.7 & 97.52 \\
1fu\_Lu4H7\_75 & 0.18 & $C2$ & 0.02 & 0.36 & -34.77 & 99.04 \\
2fu\_Lu4H7\_497 & 0.23 & $Pmn2_1$ & 0.03 & 0.34 & -35.0 & 45.59 \\
2fu\_Lu4H9\_430 & 0.13 & $Cm$ & 0.03 & 0.30 & -35.65 & 95.81 \\
1fu\_Lu4H7\_318 & 0.12 & $Amm2$ & 0.03 & 0.34 & -36.45 & 97.97 \\
1fu\_Lu4H7N\_78 & 0.24 & $P\bar{4}2m$ & 0.03 & 0.34 & -36.47 & 67.72 \\
1fu\_Lu4H7\_478 & 0.04 & $Amm2$ & 0.03 & 0.34 & -36.53 & 99.15 \\
1fu\_Lu4H7\_417 & 0.03 & $P\bar{4}3m$ & 0.02 & 0.35 & -36.77 & 97.87 \\
2fu\_Lu4H9\_46 & 0.14 & $P1$ & 0.04 & 0.27 & -36.96 & 15.41 \\
1fu\_Lu4H7\_32 & 0.16 & $Cmmm$ & 0.02 & 0.39 & -36.98 & 77.44 \\
2fu\_Lu4H7\_140 & 0.18 & $P6_3mc$ & 0.03 & 0.33 & -37.09 & 12.13 \\
2fu\_Lu4H7\_169 & 0.16 & $P\bar{1}$ & 0.02 & 0.34 & -37.64 & 83.42 \\
2fu\_Lu4H7\_375 & 0.09 & $Cmc2_1$ & 0.03 & 0.33 & -37.66 & 95.62 \\
1fu\_Lu4H9\_54 & 0.16 & $C2/m$ & 0.03 & 0.29 & -37.73 & 77.29 \\
1fu\_Lu4H11N\_161 & 0.23 & $Cm$ & 0.13 & 0.18 & -37.94 & 66.75 \\
1fu\_Lu4H7\_64 & 0.07 & $Cm$ & 0.03 & 0.31 & -37.98 & 97.45 \\
2fu\_Lu4H9\_128 & 0.23 & $C2$ & 0.04 & 0.26 & -38.11 & 45.05 \\
1fu\_Lu4H7\_288 & 0.16 & $Amm2$ & 0.02 & 0.35 & -38.25 & 65.62 \\
1fu\_Lu4H7\_118 & 0.20 & $Cm$ & 0.03 & 0.33 & -38.38 & 64.77 \\
1fu\_Lu4H7\_313 & 0.19 & $Amm2$ & 0.03 & 0.32 & -38.59 & 58.34 \\
2fu\_Lu4H7\_323 & 0.24 & $C2$ & 0.03 & 0.33 & -38.65 & 35.99 \\
1fu\_Lu4H11N\_124 & 0.22 & $C2/m$ & 0.04 & 0.26 & -38.83 & 67.27 \\
1fu\_Lu4H7\_467 & 0.23 & $P\bar{4}m2$ & 0.02 & 0.39 & -38.88 & 59.39 \\
2fu\_Lu4H9\_152 & 0.23 & $Cm$ & 0.04 & 0.27 & -38.96 & 48.52 \\
1fu\_Lu4H7\_185 & 0.16 & $Cmmm$ & 0.02 & 0.38 & -39.04 & 75.89 \\
1fu\_Lu4H7\_138 & 0.16 & $Cmmm$ & 0.02 & 0.38 & -39.04 & 76.86 \\
1fu\_Lu4H7\_96 & 0.16 & $Cmmm$ & 0.02 & 0.38 & -39.3 & 77.47 \\
1fu\_Lu4H7N\_222 & 0.10 & $Cm$ & 0.03 & 0.34 & -39.36 & 39.69 \\
2fu\_Lu4H9\_401 & 0.02 & $P4/mmm$ & 0.04 & 0.26 & -39.37 & 99.12 \\
2fu\_Lu4H7\_63 & 0.17 & $Cmc2_1$ & 0.03 & 0.32 & -39.55 & 62.64 \\
2fu\_Lu4H9\_475 & 0.14 & $Pmn2_1$ & 0.04 & 0.26 & -39.76 & 24.51 \\
2fu\_Lu4H7\_59 & 0.24 & $P1$ & 0.02 & 0.36 & -39.81 & 46.70 \\
2fu\_Lu4H9\_203 & 0.16 & $P2/c$ & nan & 0.27 & -39.93 & 52.38 \\
2fu\_Lu4H9\_192 & 0.22 & $Cm$ & 0.04 & 0.25 & -40.27 & 36.96 \\
1fu\_Lu4H7N\_405 & 0.10 & $Cm$ & 0.04 & 0.31 & -40.57 & 44.22 \\
1fu\_Lu4H7\_29 & 0.16 & $P1$ & 0.02 & 0.33 & -40.67 & 68.36 \\
2fu\_Lu4H9\_44 & 0.14 & $Cm$ & 0.04 & 0.26 & -40.71 & 72.21 \\
2fu\_Lu4H7\_58 & 0.20 & $Cc$ & 0.02 & 0.32 & -41.01 & 54.10 \\
1fu\_Lu4H7N\_358 & 0.10 & $R3m$ & 0.02 & 0.36 & -41.32 & 68.35 \\
2fu\_Lu4H9\_225 & 0.20 & $P1$ & 0.03 & 0.26 & -41.51 & 97.85 \\
3fu\_Lu2H5N\_311 & 0.19 & $P1$ & 0.04 & 0.26 & -41.68 & 44.71 \\
2fu\_Lu4H9\_138 & 0.08 & $Pmm2$ & 0.03 & 0.26 & -41.69 & 99.04 \\
1fu\_Lu4H8N\_326 & 0.18 & $P\bar{4}m2$ & 0.03 & 0.30 & -41.74 & 86.35 \\
2fu\_Lu4H7\_346 & 0.12 & $Pc$ & 0.02 & 0.32 & -41.82 & 98.34 \\
1fu\_Lu4H8N\_491 & 0.12 & $Cm$ & 0.05 & 0.26 & -41.95 & 82.29 \\
2fu\_Lu4H7\_39 & 0.12 & $Cmmm$ & 0.02 & 0.31 & -41.95 & 79.61 \\
2fu\_Lu4H9\_436 & 0.13 & $Cc$ & 0.04 & 0.25 & -42.11 & 14.54 \\
1fu\_Lu4H8N\_444 & 0.15 & $Amm2$ & 0.04 & 0.28 & -42.3 & 69.01 \\
2fu\_Lu4H9\_117 & 0.23 & $Pm$ & 0.03 & 0.26 & -42.35 & 69.30 \\
1fu\_Lu4H8N\_141 & 0.20 & $Cm$ & 0.03 & 0.31 & -42.57 & 58.82 \\
2fu\_Lu4H9\_184 & 0.10 & $I4mm$ & 0.02 & 0.28 & -42.75 & 22.18 \\
2fu\_Lu4H9\_380 & 0.10 & $Cmcm$ & 0.03 & 0.26 & -43.49 & 95.13 \\
1fu\_Lu4H7\_367 & 0.19 & $P1$ & 0.03 & 0.28 & -43.5 & 48.26 \\
2fu\_Lu4H9\_113 & 0.15 & $P1$ & 0.03 & 0.25 & -43.84 & 58.26 \\
1fu\_Lu4H9N\_216 & 0.21 & $P1$ & 0.03 & 0.26 & -44.17 & 49.05 \\
1fu\_Lu4H7\_232 & 0.14 & $R3m$ & 0.02 & 0.31 & -44.58 & 63.09 \\
1fu\_Lu4H7\_81 & 0.17 & $Cm$ & 0.02 & 0.31 & -44.59 & 57.65 \\
1fu\_Lu4H9\_167 & 0.01 & $I4/mmm$ & 0.03 & 0.25 & -44.64 & 98.97 \\
2fu\_Lu4H9\_299 & 0.01 & $I4/mmm$ & 0.03 & 0.25 & -44.64 & 98.92 \\
1fu\_Lu4H7\_386 & 0.19 & $Pmm2$ & 0.02 & 0.32 & -44.88 & 60.30 \\
1fu\_Lu4H8N\_5 & 0.07 & $R\bar{3}m$ & 0.03 & 0.27 & -44.97 & 51.47 \\
1fu\_Lu4H7N\_471 & 0.15 & $Pmm2$ & 0.03 & 0.28 & -45.06 & 67.44 \\
2fu\_Lu4H9\_191 & 0.15 & $Cmm2$ & 0.04 & 0.23 & -45.06 & 76.30 \\
1fu\_Lu4H7\_44 & 0.19 & $C2$ & 0.02 & 0.30 & -45.25 & 58.91 \\
2fu\_Lu4H9\_210 & 0.24 & $P2_1$ & 0.03 & 0.25 & -45.38 & 84.27 \\
2fu\_Lu4H9\_111 & 0.10 & $Cc$ & 0.03 & 0.24 & -45.43 & 91.48 \\
1fu\_Lu4H7N\_446 & 0.23 & $Cm$ & 0.05 & 0.25 & -45.93 & 74.54 \\
1fu\_Lu4H7N\_166 & 0.15 & $P1$ & 0.04 & 0.26 & -45.95 & 27.14 \\
2fu\_Lu3H8N\_4 & 0.23 & $P1$ & 0.04 & 0.22 & -45.95 & 71.09 \\
1fu\_Lu4H9\_235 & 0.15 & $Cmm2$ & 0.03 & 0.24 & -45.96 & 73.67 \\
2fu\_Lu4H7\_456 & 0.22 & $Cm$ & 0.02 & 0.28 & -46.0 & 58.31 \\
2fu\_Lu2H5N\_142 & 0.23 & $Pma2$ & 0.20 & 0.14 & -46.34 & 8.76 \\
1fu\_Lu4H9\_194 & 0.11 & $R\bar{3}m$ & 0.02 & 0.26 & -46.36 & 91.31 \\
1fu\_Lu4H11N\_150 & 0.16 & $P1$ & 0.03 & 0.25 & -46.41 & 49.40 \\
1fu\_Lu4H9N\_10 & 0.21 & $P3m1$ & 0.04 & 0.24 & -46.57 & 37.17 \\
1fu\_Lu4H9\_6 & 0.16 & $P1$ & 0.04 & 0.22 & -46.88 & 64.86 \\
2fu\_Lu3H8N\_116 & 0.20 & $P1$ & 0.03 & 0.24 & -47.27 & 49.77 \\
2fu\_Lu4H9\_336 & 0.22 & $Pmc2_1$ & 0.04 & 0.22 & -47.54 & 74.96 \\
1fu\_Lu4H7\_223 & 0.20 & $P2$ & 0.02 & 0.28 & -47.64 & 73.17 \\
2fu\_Lu4H9\_364 & 0.24 & $Cc$ & 0.04 & 0.21 & -48.79 & 94.71 \\
1fu\_Lu4H9N\_136 & 0.04 & $P3m1$ & 0.31 & 0.11 & -48.99 & 13.12 \\
1fu\_Lu4H7\_340 & 0.10 & $Pmmm$ & 0.02 & 0.30 & -49.15 & 98.19 \\
2fu\_Lu4H9\_323 & 0.19 & $C2/m$ & 0.03 & 0.23 & -49.35 & 53.14 \\
1fu\_Lu4H11N\_251 & 0.19 & $Cm$ & 0.05 & 0.19 & -49.69 & 66.42 \\
1fu\_Lu4H8N\_489 & 0.23 & $Cm$ & 0.03 & 0.24 & -50.02 & 68.51 \\
1fu\_Lu4H7\_378 & 0.14 & $Cmm2$ & 0.02 & 0.27 & -50.04 & 79.79 \\
1fu\_Lu4H7\_325 & 0.24 & $Amm2$ & 0.02 & 0.25 & -50.53 & 71.05 \\
1fu\_Lu4H8N\_436 & 0.16 & $P\bar{4}$ & 0.04 & 0.22 & -50.6 & 11.86 \\
1fu\_Lu2H5N\_194 & 0.17 & $Cm$ & 0.39 & 0.10 & -50.89 & 62.99 \\
2fu\_Lu4H9\_335 & 0.22 & $P4/m$ & 0.02 & 0.23 & -51.35 & 13.51 \\
1fu\_Lu4H7\_307 & 0.12 & $P2$ & 0.02 & 0.25 & -51.75 & 83.52 \\
2fu\_Lu4H11N\_548 & 0.22 & $P1$ & 0.03 & 0.21 & -51.94 & 73.65 \\
1fu\_Lu4H7\_228 & 0.21 & $Pm$ & 0.02 & 0.25 & -51.95 & 52.84 \\
2fu\_Lu4H11N\_595 & 0.16 & $P1$ & 0.04 & 0.19 & -52.22 & 55.28 \\
1fu\_Lu4H7N\_375 & 0.17 & $Cm$ & 0.02 & 0.27 & -52.44 & 42.71 \\
1fu\_Lu4H7N\_280 & 0.19 & $P1$ & 0.03 & 0.23 & -52.51 & 42.08 \\
1fu\_Lu4H7N\_325 & 0.20 & $Amm2$ & 0.04 & 0.22 & -52.52 & 96.86 \\
1fu\_Lu4H7N\_11 & 0.11 & $Cm$ & 0.03 & 0.24 & -52.56 & 91.97 \\
1fu\_Lu4H8N\_308 & 0.15 & $Pm$ & 0.03 & 0.22 & -52.9 & 26.63 \\
1fu\_Lu4H8N\_9 & 0.21 & $P4mm$ & 0.04 & 0.21 & -53.22 & 64.27 \\
2fu\_Lu3H8N\_186 & 0.22 & $P6_3/mmc$ & 0.06 & 0.16 & -53.34 & 65.81 \\
1fu\_Lu4H7\_455 & 0.15 & $Pmm2$ & 0.02 & 0.23 & -53.65 & 94.68 \\
1fu\_Lu4H8N\_494 & 0.14 & $Amm2$ & 0.03 & 0.23 & -53.77 & 28.24 \\
2fu\_Lu4H9\_332 & 0.23 & $C2$ & 0.03 & 0.19 & -53.89 & 60.13 \\
1fu\_Lu4H9N\_385 & 0.17 & $Pm$ & 0.04 & 0.19 & -53.99 & 71.40 \\
1fu\_Lu4H8N\_198 & 0.10 & $Pmm2$ & 0.03 & 0.22 & -54.2 & 72.35 \\
1fu\_Lu4H9\_216 & 0.15 & $P\bar{6}m2$ & 0.02 & 0.21 & -55.28 & 42.84 \\
1fu\_Lu4H7\_379 & 0.20 & $Pm$ & 0.02 & 0.22 & -55.31 & 88.22 \\
2fu\_Lu4H9\_444 & 0.24 & $P2_1/m$ & 0.03 & 0.19 & -55.58 & 60.56 \\
1fu\_Lu4H8N\_410 & 0.14 & $Cm$ & 0.04 & 0.19 & -55.64 & 76.18 \\
1fu\_Lu4H9\_52 & 0.16 & $Amm2$ & 0.02 & 0.20 & -55.91 & 60.98 \\
1fu\_Lu4H8N\_105 & 0.07 & $Pm$ & 0.03 & 0.21 & -56.18 & 74.46 \\
1fu\_Lu4H8N\_46 & 0.14 & $Pm$ & 0.03 & 0.20 & -56.19 & 27.61 \\
1fu\_Lu4H7N\_377 & 0.15 & $Cm$ & 0.03 & 0.20 & -56.47 & 21.50 \\
3fu\_Lu2H5N\_89 & 0.23 & $C2$ & 0.03 & 0.20 & -56.82 & 57.28 \\
1fu\_Lu4H9N\_83 & 0.19 & $Pm$ & 0.04 & 0.17 & -56.92 & 81.91 \\
1fu\_Lu4H7N\_491 & 0.14 & $P1$ & 0.03 & 0.21 & -57.5 & 84.14 \\
1fu\_Lu4H9\_129 & 0.15 & $Imm2$ & 0.03 & 0.17 & -57.64 & 14.80 \\
1fu\_Lu4H7N\_172 & 0.17 & $P1$ & 0.03 & 0.21 & -57.7 & 83.10 \\
2fu\_Lu4H9\_92 & 0.22 & $Cc$ & 0.03 & 0.17 & -57.74 & 45.77 \\
2fu\_Lu4H7N\_206 & 0.19 & $P4_2mc$ & 0.03 & 0.21 & -57.84 & 27.93 \\
1fu\_Lu4H7N\_48 & 0.23 & $Pm$ & 0.03 & 0.19 & -58.03 & 63.99 \\
2fu\_Lu4H9\_67 & 0.14 & $P4_2/nmc$ & 0.02 & 0.20 & -58.06 & 40.45 \\
2fu\_Lu4H9\_115 & 0.19 & $C2/c$ & 0.04 & 0.15 & -58.36 & 33.31 \\
1fu\_Lu4H7N\_488 & 0.17 & $Cm$ & 0.02 & 0.22 & -58.37 & 67.76 \\
1fu\_Lu4H7N\_49 & 0.10 & $I\bar{4}2m$ & 0.03 & 0.20 & -58.61 & 59.77 \\
1fu\_Lu4H8N\_334 & 0.08 & $Cm$ & 0.03 & 0.18 & -58.69 & 68.93 \\
1fu\_Lu4H8N\_72 & 0.22 & $Pm$ & 0.03 & 0.18 & -58.99 & 64.08 \\
4fu\_Lu2H5N\_206 & 0.20 & $P1$ & 0.03 & 0.17 & -59.1 & 61.21 \\
1fu\_Lu4H8N\_150 & 0.14 & $Cm$ & 0.02 & 0.19 & -59.37 & 32.61 \\
1fu\_Lu4H7N\_420 & 0.13 & $Pmm2$ & 0.03 & 0.19 & -59.45 & 65.32 \\
1fu\_Lu4H8N\_37 & 0.16 & $Cmmm$ & 0.02 & 0.21 & -59.55 & 65.12 \\
1fu\_Lu4H8N\_224 & 0.07 & $Amm2$ & 0.03 & 0.18 & -59.69 & 46.12 \\
1fu\_Lu3H8N\_417 & 0.23 & $Cm$ & 0.04 & 0.14 & -60.17 & 57.58 \\
1fu\_Lu4H8N\_196 & 0.18 & $Cm$ & 0.03 & 0.18 & -60.58 & 14.69 \\
1fu\_Lu4H7N\_107 & 0.21 & $C2$ & 0.03 & 0.18 & -60.64 & 66.81 \\
1fu\_Lu4H7N\_372 & 0.15 & $P4mm$ & 0.03 & 0.19 & -60.64 & 52.78 \\
1fu\_Lu4H8N\_216 & 0.16 & $Cm$ & 0.03 & 0.17 & -61.08 & 60.05 \\
2fu\_Lu4H8N\_326 & 0.10 & $P4_2/nmc$ & 0.02 & 0.18 & -61.3 & 67.61 \\
1fu\_Lu4H8N\_119 & 0.13 & $P1$ & 0.03 & 0.17 & -61.35 & 64.41 \\
1fu\_Lu4H7N\_424 & 0.09 & $P\bar{4}3m$ & 0.02 & 0.19 & -61.7 & 61.73 \\
2fu\_Lu4H9\_420 & 0.04 & $Cmcm$ & 0.03 & 0.15 & -62.03 & 93.12 \\
3fu\_Lu2H5N\_336 & 0.14 & $P\bar{3}m1$ & 0.03 & 0.16 & -62.78 & 46.56 \\
1fu\_Lu4H7N\_293 & 0.22 & $C2$ & 0.03 & 0.16 & -63.08 & 49.69 \\
1fu\_Lu4H7N\_190 & 0.21 & $Cm$ & 0.02 & 0.17 & -63.73 & 65.93 \\
1fu\_Lu4H8N\_83 & 0.18 & $Cm$ & 0.03 & 0.15 & -64.17 & 57.73 \\
1fu\_Lu4H7N\_494 & 0.15 & $Cm$ & 0.02 & 0.16 & -64.6 & 63.68 \\
2fu\_Lu2H5N\_194 & 0.08 & $P1$ & 0.02 & 0.14 & -66.45 & 63.50 \\
1fu\_Lu4H9N\_167 & 0.22 & $R3m$ & 0.03 & 0.13 & -66.49 & 42.43 \\
1fu\_Lu4H7N\_495 & 0.22 & $Cm$ & 0.02 & 0.14 & -67.18 & 23.42 \\
1fu\_Lu4H7N\_67 & 0.02 & $R\bar{3}m$ & 0.03 & 0.06 & -76.61 & 53.46 \\
1fu\_Lu2H5N\_439 & 0.21 & $Cm$ & 0.28 & \textit{N/A} & \textit{N/A} & 61.86 \\
1fu\_Lu3H8N\_431 & 0.13 & $Cm$ & 0.04 & \textit{N/A} & \textit{N/A} & 86.07 \\
1fu\_Lu4H11N\_196 & 0.17 & $C2$ & 0.39 & \textit{N/A} & \textit{N/A} & 60.82 \\
1fu\_Lu4H8N\_175 & 0.13 & $C2$ & 0.03 & \textit{N/A} & \textit{N/A} & 81.56 \\
1fu\_Lu4H9\_127 & 0.20 & $Cm$ & 0.03 & \textit{N/A} & \textit{N/A} & 61.62 \\
1fu\_Lu4H9\_74 & 0.01 & $C2/m$ & 0.04 & \textit{N/A} & \textit{N/A} & 99.19 \\
2fu\_Lu4H9\_63 & 0.02 & $I4/mmm$ & 0.04 & \textit{N/A} &\textit{N/A} & 99.24 \\

\end{longtable}
\end{center}

\clearpage

\section{Bader charge analysis}
\begin{figure}[ht]
\includegraphics[angle=0,width=0.99\textwidth]{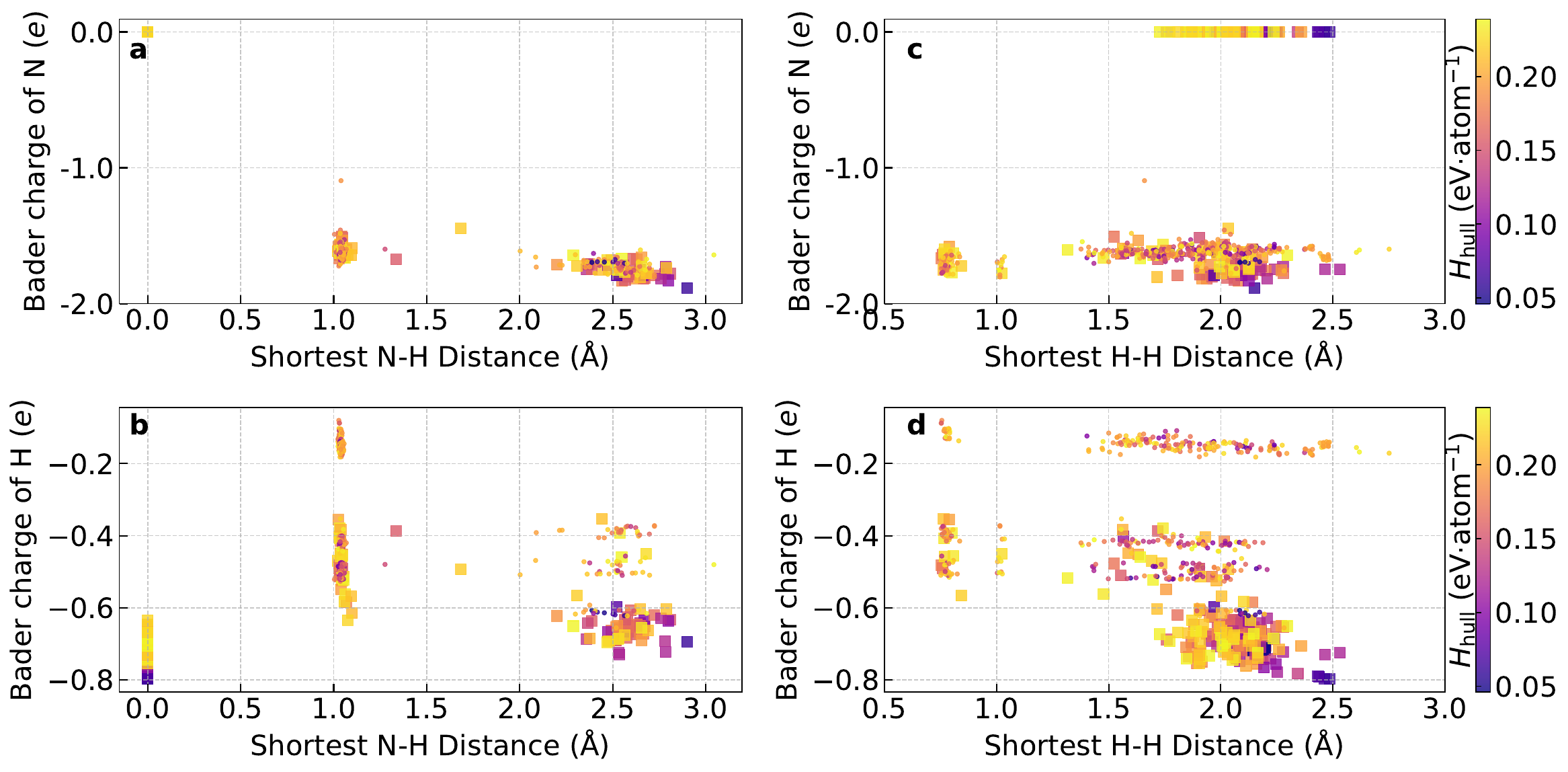}
\caption{\label{fig:bader-charge}  
\textbf{The Bader charge of N and H versus the shortest N-H and H-H distances.} 
\textbf{a} The Bader charge of N versus the shortest N-H distance. \textbf{b} The Bader charge of H versus the shortest N-H distance. 
\textbf{c} The Bader charge of N versus the shortest H-H distance. \textbf{d} The Bader charge of H versus the shortest H-H distance. 
The color bar shows the enthalpy distance above the convex hull. The square markers and circle markers represent metallic states and insulating states, respectively. Note that the Bader charge of N is set to zero artificially for all binary compounds.
}
\end{figure}

\section{The analysis about H sites and Lu sublattice}
\begin{figure}[ht]
\includegraphics[angle=0,width=0.99\textwidth]{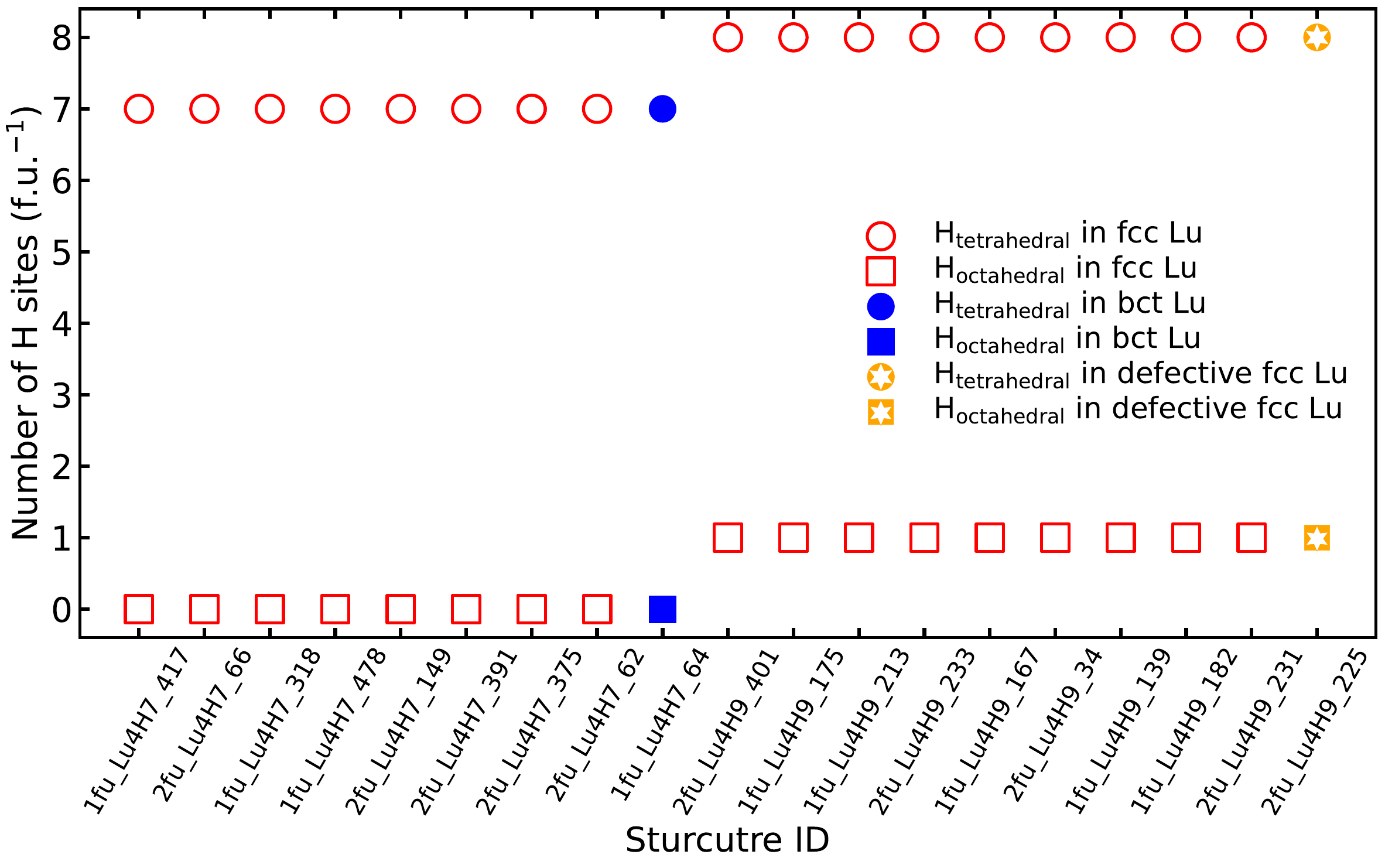}
\caption{\label{fig:H_sites_Lu_lattice}  
\textbf{Analysis of hydrogen sites and lutetium sublattice.} H$_{\rm tetrahedral}$ and H$_{\rm octahedral}$ represent the hydrogen atoms at the interstitial tetrahedral sites and interstitial octahedral sites, respectively. The abbreviations fcc and bct stand for face-centered cubic and body-centered tetragonal, respectively. 
}
\end{figure}

\begin{figure}
    \centering
    \includegraphics[width=0.99\textwidth]{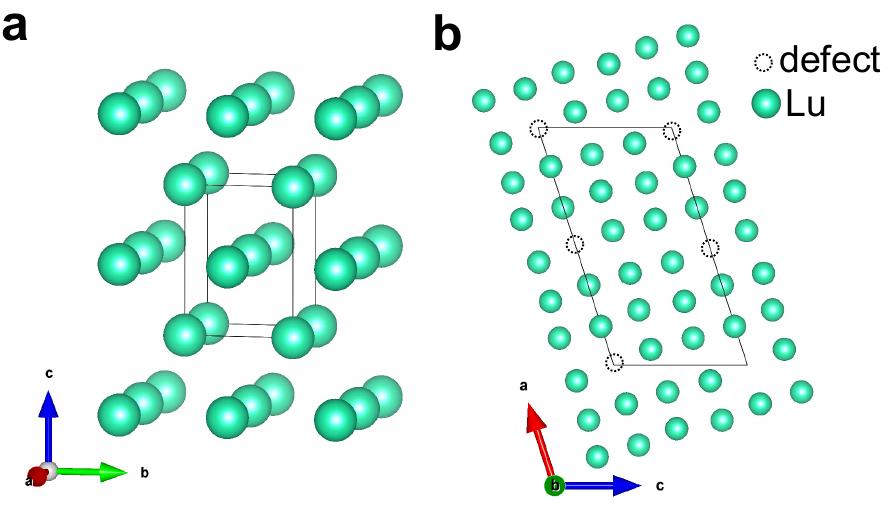}
    \caption{The two distinct Lu sublattices. \textbf{a}  Body-centered tetragonal (bct) configuration of Lu sublattice in 1fu\_Lu4H7\_64. \textbf{b} Defective face-centered cubic Lu sublattice in 2fu\_Lu4H9\_225. The dashed circles indicate the Lu defects. The solid lines indicate the primitive cell of the Lu sublattices. The H and N atoms are removed for clarity of the Lu sublattice.}
    \label{fig:Lu-sublattice}
\end{figure}

Fig.~\ref{fig:Lu-sublattice} shows the body-centered tetragonal (bct) Lu sublattice in 1fu\_Lu4H7\_64 and defective face-centered cubic (fcc) Lu sublattice 2fu\_Lu4H9\_225. In 2fu\_Lu4H9\_225, the restoration of Lu atoms in the defective sites will lead to the return of the Lu sublattice to fcc structure. This is why this Lu sublattice of 2fu\_Lu4H9\_225 is named as ``defective fcc'' structure in our study.

\section{Comparison between ${Fm\bar{3}m}$ LuH$_3$  and ${Pm\bar{3}m}$ Lu$_4$H$_{11}$N}

\begin{figure}
    \centering
    \includegraphics[width=0.99\textwidth]{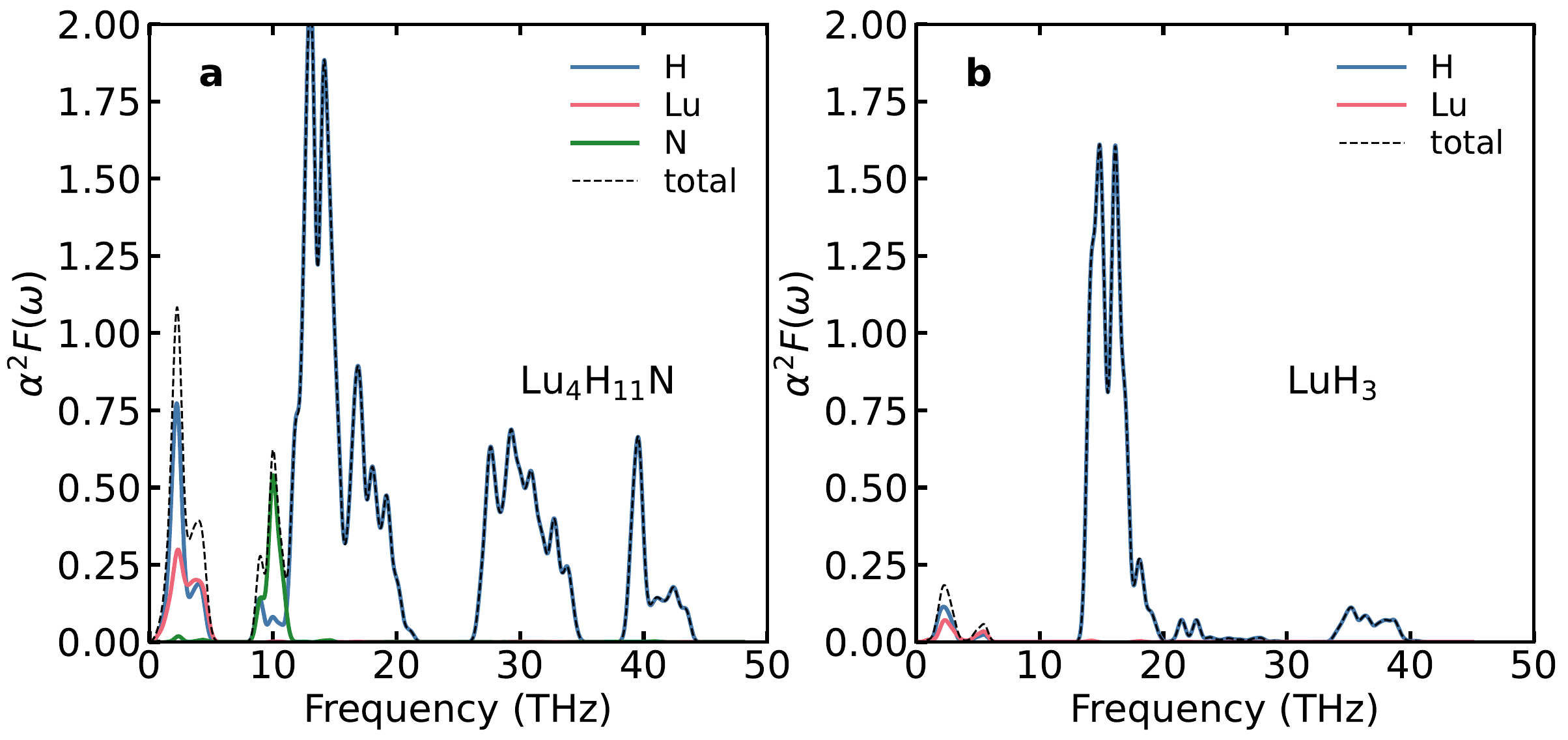}
    \caption{Element projected Eliashberg spectral functions of (\textbf{a}) Lu$_4$H$_{11}$N and (\textbf{b}) LuH$_3$ at 20 GPa and 300 K.}
    \label{fig:comp_a2f}
\end{figure}

In Fig.~\ref{fig:comp_a2f} we show element projected Eliashberg spectral functions of ${Pm\bar{3}m}$ Lu$_4$H$_{11}$N and LuH$_3$ at 20 GPa. In both cases, most of the electron-phonon coupling comes from H-dominated modes, with small contributions from Lu atoms for the low-frequency region and, in the case of Lu$_4$H$_{11}$N, N-dominated modes in the region around 10 THz. From this, we can see that N doping has an indirect effect through the displacement of H atoms from perfect tetrahedral positions which leads to their stronger coupling to electrons. Additionally, the electronic DOS at the Fermi level increases from 2.43131 Ry$^{-1}$ per spin per LuH$_3$ in the case of pure LuH$_3$ to 4.702 Ry$^{-1}$ per spin per LuH$_3$ in the doped case. These two effects explain the threefold increase in estimated critical temperature from pure LuH$_3$ to Lu$_4$H$_{11}$N. However, here we would like to point out again that doping N in LuH$_3$ pushes it further off the convex hull (from 82 meV/atom for LuH$_3$ to 270 meV/atom for Lu$_4$H$_{11}$N), and that the superconducting properties are determined primarily by the hydrogen atoms.

\begin{figure}
    \centering
    \includegraphics[width=0.99\textwidth]{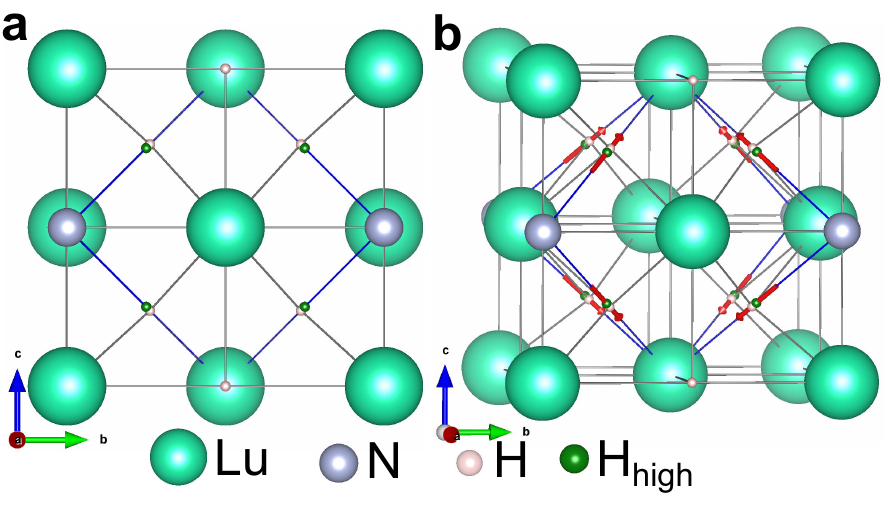}
    \caption{The crystal structure of ${Pm\bar{3}m}$ Lu$_4$H$_{11}$N at 20 GPa depicted from two different perspectives. 
    H$_{\rm high}$ (green sphere) refers to the original H atoms at the high-symmetry sites before N is introduced. The red arrows indicate the displacements of hydrogen atoms at the tetrahedral sites resulting from the introduction of N. The diagonal blue lines are drawn to connect nitrogen and lutetium atoms, serving as visual guides}
    \label{fig:comparison-LuH3-Lu4H11N}
\end{figure}

Table~\ref{table:comparison-LuH3-Lu4H11N} displays the lattice parameters and atomic positions of ${Fm\bar{3}m}$ LuH$_3$  and ${Pm\bar{3}m}$ Lu$_4$H$_{11}$N. The tetrahedral hydrogen sites that originally have the high-symmetry Wyckoff sites (0.25, 0.25, 0.25) are displaced to (0.23835, 0.23835, 0.23835) due to the doping of N at the octahedral site. This change in structure provides evidence that the presence of nitrogen reduces the symmetry of hydrogen sites at the tetrahedral positions. The displacements of the hydrogen sites resulting from the introduction of nitrogen are schematically shown in Fig.~\ref{fig:comparison-LuH3-Lu4H11N}.

\begin{table}[]
\caption{\label{table:comparison-LuH3-Lu4H11N}
The comparison of lattice parameters and atomic positions of ${Fm\bar{3}m}$ LuH$_3$  and ${Pm\bar{3}m}$ Lu$_4$H$_{11}$N at 20 GPa. The Wyckoff sites are given in fractional coordinates.} 
\begin{tabular}{cclcccc}
\hline
\hline
\multicolumn{4}{c}{LuH$_3$ ($a$ = 4.82224 \AA)}                                            & \multicolumn{3}{c}{Lu$_4$H$_{11}$N ($a$ = 4.84550 \AA)}                               \\ \hline
Atom site & \multicolumn{2}{c}{Wyckoff labels} & Wyckoff sites      & Atom symbol & Wyckoff label & Wyckoff sites               \\
Lu1       & \multicolumn{2}{c}{4a}             & 0, 0, 0            & Lu1         & 3c            & (0, 0.5, 0.5)               \\
H1        & \multicolumn{2}{c}{8c}             & (0.25, 0.25, 0.25) & Lu2         & 1a            & (0, 0, 0)                   \\
H2        & \multicolumn{2}{c}{4b}             & (0.5, 0.5, 0.5)    & H1          & 8g            & (0.23835, 0.23835, 0.23835) \\ \cline{1-4}\cline{1-4}
\multicolumn{4}{c}{\multirow{2}{*}{}}                               & H2          & 3d            & (0.5, 0, 0)                 \\
\multicolumn{4}{c}{}                                                & N1          & 1b            & (0.5, 0.5, 0.5)             \\ \cline{5-7} \cline{5-7}
\end{tabular}
\end{table}

\clearpage
\newpage

%
\end{document}